\documentclass[AMA,STIX1COL]{WileyNJD-v2}
\articletype{Research Article}%

\received{xxxx}
\revised{xxxx}
\accepted{xxxx}
\usepackage{etoolbox}
\makeatletter
\patchcmd{\@sect}{\uppercase{#8}}{#8}{}{}
\usepackage{comment}
\usepackage{rotating}
\usepackage{cases}
\usepackage{threeparttable}
\usepackage{subcaption}
\usepackage{placeins}
\usepackage{graphicx}
\usepackage{rotating}
\newcommand{\be}{\begin{eqnarray}}
\newcommand{\ee}{\end{eqnarray}}
\newcommand{\bee}{\begin{eqnarray*}}
\newcommand{\eee}{\end{eqnarray*}}
\newcommand{\bi}{\begin{enumerate}}
\newcommand{\ei}{\end{enumerate}}

\begin{document}

\title{Which Small-Sample Correction Should Be Used When Analyzing Stepped-Wedge Designs with Time-Varying Treatment Effects?}

\author[1,2]{Yongdong Ouyang}
\author[3,4]{Monica Taljaard}
\author[5]{James P. Hughes$^{\dagger}$}
\author[6,7]{Fan Li$^{\dagger}$}

\begingroup
\renewcommand{\thefootnote}{}
\footnotetext{$^{\dagger}$ James P. Hughes and Fan Li jointly supervised this work.}
\endgroup

\authormark{Ouyang et al.}

\address[1]{\orgdiv{Department of Biostatistics and Bioinformatics}, \orgname{Roswell Park Comprehensive Cancer Center}, \orgaddress{\state{Buffalo, New York}, \country{USA}}}
\address[2]{\orgdiv{Roswell Park Graduate Division}, \orgname{State University of New York at Buffalo}, \orgaddress{\state{Buffalo, New York}, \country{USA}}}
\address[3]{\orgdiv{Methodological and Implementation Research}, \orgname{Ottawa Hospital Research Institute}, \orgaddress{\state{Ottawa, Ontario}, \country{Canada}}}
\address[4]{\orgdiv{School of Epidemiology and Public Health}, \orgname{University of Ottawa}, \orgaddress{\state{Ottawa, Ontario}, \country{Canada}}}
\address[5]{\orgdiv{Department of Biostatistics, School of Public Health}, \orgname{University of Washington}, \orgaddress{\state{Seattle, Washington}, \country{USA}}}
\address[6]{\orgdiv{Department of Biostatistics}, \orgname{Yale University}, \orgaddress{\state{New Haven, Connecticut}, \country{USA}}}
\address[7]{\orgdiv{Center for Methods in Implementation and Prevention Science}, \orgname{Yale University}, \orgaddress{\state{New Haven, Connecticut}, \country{USA}}}

\corres{Yongdong Ouyang, Department of Biostatistics and Bioinformatics, Roswell Park Comprehensive Cancer Center, Buffalo, New York, USA,\\
 \email{youyang2@buffalo.edu}}

\abstract[Summary]{Stepped-wedge cluster randomized trials (SW-CRTs) evaluate interventions rolled out across clusters over time. Standard analyses typically use immediate-treatment (IT) models, which assume effects begin at crossover and remain constant thereafter. When effects vary with exposure duration, IT models may misrepresent target effects. Exposure-time indicator (ETI) models address this by allowing treatment effects to differ by time since exposure and by targeting the time-averaged treatment effect (TATE) and long-term effect (LTE). Like IT models, ETI models require specification of a random-effects structure, which is often misspecified, and the performance of robust variance estimators (RVEs) in this setting is not well understood. We review RVEs for ETI models and evaluate them in simulation studies with continuous and binary outcomes under correctly specified (binary only) and misspecified random-effects structures. We compare the classic sandwich, Kauermann–Carroll (KC), Mancl–DeRouen (MD), and Morel–Bokossa–Neerchal (MBN) estimators for inference on the TATE and LTE. Our simulations show that under misspecified random-effects structures, model-based standard errors (SE) produced undercoverage, whereas RVEs improved performance. For continuous outcomes, MD with a $t-$distribution and degrees of freedom equal to the number of clusters minus two gave the most consistent coverage probabilities. For binary outcomes, MBN was the only consistently reliable option. MD, however, could be unstable in one-cluster-per-sequence designs because of data sparsity. Across scenarios, both model-based SE and RVE for LTE were unstable, indicating that greater caution is needed when targeting LTE under ETI models.}



\maketitle

\section{Introduction}
\begin{figure}[!htbp]
    \centering
    \includegraphics[width=0.70\linewidth]{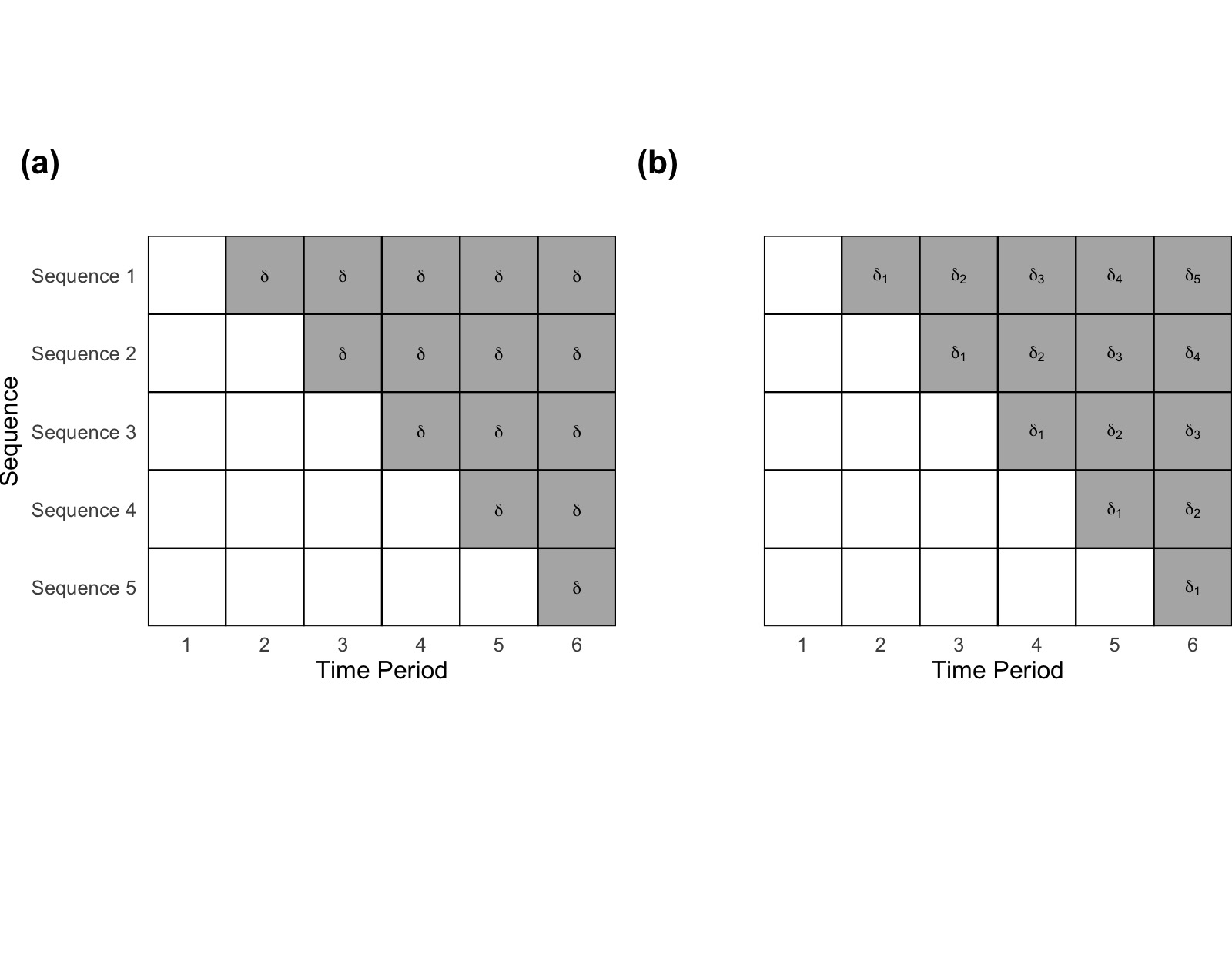}
    \caption{Both diagrams illustrate a standard stepped-wedge cluster randomized trial with 5 sequences and 6 periods. Dark shading indicates intervention periods, and light shading indicates control periods. Panel (a) shows the immediate-treatment (IT) effect model, which assumes that the treatment effect begins at crossover and remains constant thereafter. Panel (b) shows the exposure-time indicator (ETI) effect model, in which the treatment effect depends on exposure time \(e\) since crossover. \(\delta_e\) denotes the treatment effect at exposure time \(e\).}
    \label{fig:swdeisgn}
\end{figure}

Cluster randomized trials (CRTs) involve randomizing groups rather than individuals to evaluate interventions and are a common and useful design when there is a risk of contamination or the intervention must be delivered at the group level.\cite{donner_design_2000} Due to shared characteristics, individuals within the same cluster tend to be more similar than individuals from different clusters. This within-cluster similarity is commonly measured by the intra-cluster correlation coefficient (ICC).\cite{donner_design_2000,ouyang_estimating_2023} An increasingly popular variant of CRTs is the stepped-wedge cluster randomized trial (SW-CRT), which incorporates a longitudinal component by taking repeated measurements from clusters over time \cite{hussey_design_2007}. In a standard SW-CRT (Figure~\ref{fig:swdeisgn}a), all clusters begin in the control condition and subsequently transition (in a randomized fashion) to the intervention condition at different prespecified time points. By the final period, all clusters are exposed to the intervention by design. Depending on the type of design, outcomes can be collected from different individuals in each period (a cross-sectional design), the same individuals (a closed cohort design) over multiple periods, or a mix of both (an open cohort design) \cite{copas_designing_2015}. 

\FloatBarrier

Mixed‐effects models are among the most widely used and practically accessible analytic tools for SW‐CRTs, as they flexibly represent hierarchical data structures and account for within‐cluster correlation through random effects \cite{li_mixed-effects_2021}. Compared with parallel‐arm CRTs, SW‐CRTs usually require more complex correlation structures because observations are correlated both within clusters and across time periods. Consequently, the choice of random‐effects dictates the implied correlations among observations within periods and over time  and has direct implications for the validity and efficiency of statistical inference \cite{ouyang_maintaining_2024}. A substantial body of methodological work has therefore emerged on correlation structures tailored to SW‐CRTs, including the exchangeable (EXCH) \cite{hussey_design_2007}, nested exchangeable (NE) \cite{hooper_sample_2016,girling_statistical_2016}, and exponential‐decay (ED) models \cite{kasza_impact_2019}, which differ in how they characterize within‐period and between‐period correlations (e.g., random-effects structure) over time. Additional random effects, such as random cluster‐by‐intervention interactions, can further be incorporated to accommodate heterogeneity in treatment effects across clusters \cite{voldal_model_2022}. Although the choice of random‐effects structure can influence sample size considerations \cite{ouyang_sample_2022,ouyang_accounting_2023,kasza_impact_2019}, recent work by Ouyang et al.\cite{ouyang_maintaining_2024} shows that, at least in linear mixed models, misspecification of the random‐effects structure can be effectively mitigated through the use of robust sandwich variance estimators (RVEs), thereby preserving valid inference even under correlation misspecification. 

Just as the random-effects structure defines the assumed correlation, the fixed-effects structure defines the assumed nature of the intervention effect. The standard mixed-effects model of SW-CRTs assumes a time-invariant, immediate treatment (IT) effect (as shown in Figure~\ref{fig:swdeisgn}a~) \cite{hussey_design_2007,hemming_analysis_2017}. That is, once a cluster switches to the intervention, the model assumes that the effect begins immediately at that transition and remains constant thereafter. In practice, however, intervention effects may accumulate, decay, or emerge only after a delay \cite{hughes_current_2015, kenny_analysis_2022}. To reflect such dynamics, the exposure-time indicator (ETI) model represents the treatment effect using a set of binary indicators that encode both treatment status and time since exposure (Figure~\ref{fig:swdeisgn}b), allowing effects to vary across periods and enabling a richer interpretation of the intervention trajectory \cite{hughes_current_2015}. In SW-CRTs, clusters switch from control to intervention at different calendar times, so treatment is partially confounded with time. Crucially, clusters accumulate exposure time under the intervention. If the intervention’s impact ramps up, is delayed, or decays (e.g., a learning curve/implementation maturation), the common immediate, constant effect mixed-model analysis gave misleading results. Kenny et al. \cite{kenny_analysis_2022} show that, when treatment effects vary with exposure time, the standard IT model estimates a weighted average of exposure-time specific effects with potentially negative weights. Therefore, it can converge to the wrong magnitude and even the opposite sign of target summaries such as the time-averaged or long-term effect. Maleyeff et al. \cite{maleyeff_assessing_2023} emphasize that asking whether effects differ by exposure time is often scientifically central in SW-CRTs, and they propose more efficient ways to model exposure-time heterogeneity than rigid parametric curves or highly parameterized categorical-by-exposure approaches. Finally, Wang et al. \cite{wang_how_2024} stress that “model-robust” inference in SW-CRTs generally hinges on correct specification of  the treatment-effect structure (e.g., allowing dependence on calendar and/or exposure time), with valid estimates of uncertainty obtained via sandwich-based methods (and an added g-computation step for non-identity links/ratio estimands).

Regardless of which model we choose, one of the biggest challenges for mixed-effects models in SW-CRTs is the potential for misspecified random-effects structure, which is often exacerbated by a small number of clusters \cite{ouyang_accounting_2023,nevins_adherence_2024,tong_review_2025}. Although model-selection criteria are available, their performance can be unreliable in small samples, and the consequences of post-selection inference remain insufficiently understood \cite{rezaei-darzi_inference_2025}. RVEs offer a potential safeguard against such random-effects misspecification by providing valid standard errors (SEs) even when the working correlation is incorrect \cite{liang_longitudinal_1986}. Recent simulation evidence for continuous outcomes suggests that, within LMM analyses of SW-CRTs, a cluster-robust variance with small-sample \(t\)-based inference (notably approximate jackknife estimator coupled with a number of cluster minus two degrees-of-freedom correction) can substantially improve coverage under random-effects misspecification \cite{ouyang_maintaining_2024}. Complementarily, when the number of clusters is very small, fixed-effects analyses that condition on cluster indicators have been reported to yield more stable Type I error and coverage than mixed-effects models even when using commonly recommended Kenward-Roger or Satterthwaite corrections \cite{lee_fixed-effects_2024}. In the alternative generalized estimating equation (GEE) framework to analyze SW-CRTs, several finite-sample corrections to the sandwich variance estimator and corresponding \(t\)-based inference have been proposed for stepped-wedge settings. In particular, finite-sample adjusted sandwich estimators such as Fay-Graubard (FG) corrections have been recommended for GEE analyses of stepped-wedge trials \cite{scott_finite-sample_2017}, and broader comparisons for binary outcomes suggest that Kauermann-Carroll (KC) and Fay-Graubard corrections, paired with \(t\)-based degrees-of-freedom estimators (e.g., DF$_{FG}$), are among the most reliable options when the number of clusters is limited \cite{ford_maintaining_2020,thompson_comparison_2021}. Nonetheless, performance remains sensitive to design features and the working correlation model, and correlation misspecification can still inflate Type I error \cite{rezaei-darzi_inference_2025}. Complementary design-based approaches, including permutation or randomization-based inference using within-period (``vertical'') contrasts, have also been proposed to mitigate reliance on parametric correlation assumptions in small-sample SW-CRTs \cite{thompson_robust_2018,rabideau_randomization-based_2021}. However, almost all existing small-sample evaluations have largely focused on constant or immediate intervention effects, leaving unanswered questions about small-sample inference for ETI specifications where treatment effects evolve over exposure time. To our best knowledge, no studies have yet evaluated the performance of RVEs, including their small-sample corrections, within the ETI modeling framework for binary data. This lack of foundational evidence motivates the comprehensive evaluation of RVEs for both continuous and binary outcomes under ETI-based mixed-effects analyses.

This manuscript addresses these gaps through a comprehensive simulation study evaluating the performance of RVEs for ETI models. Within the generalized linear mixed-effects model (GLMM) framework for continuous and binary outcomes, this study systematically investigates several key questions. First, for continuous outcomes, we evaluate whether RVEs provide valid inference for time-varying treatment effects when the random-effects structure is misspecified. Second, for binary outcomes, since prior simulation studies on SW-CRTs with binary endpoints have largely focused on IT models and assumed a correctly specified random-effects structure \cite{ford_maintaining_2020,qu_small_1994,thompson_comparison_2021}, we will first establish a crucial performance baseline by evaluating RVEs for ETI models under correctly specified random-effects structure. We then examine the performance of RVEs for ETI models under a misspecified random-effects structure. Ultimately, a primary objective in both settings is to determine the most reliable RVE and small-sample correction method, particularly for use in designs with a limited number of clusters.

The remainder of this manuscript is organized as follows. Section 2 reviews the statistical models, covering both IT and ETI models with different random-effects structures. Section 3 then introduces the specific RVEs that are formally compared in our study. The design of our comprehensive simulation is detailed in Section 4. We assess RVE performance for continuous outcomes under random-effects misspecification, and for binary outcomes, we evaluate performance under both correctly specified and misspecified random-effects structures. The results of these simulations are presented in Section 5. We then present real trial example in Section 6. Section 7 concludes the manuscript with a discussion of our findings, practical recommendations, and directions for future research.

\section{Statistical models for analyzing stepped-wedge designs}
In this manuscript, we assume a GLMM framework to accommodate both continuous and binary outcomes. A GLMM links the expected value of the outcome to a linear predictor via a link function, $g(\cdot)$. The general form of this model is:
\begin{align}
g(E[Y_{ijk}]) = \eta_{ijk} \nonumber
\end{align}
where $Y_{ijk}$ is the outcome for the $k$-th individual in the $i$-th cluster during the $j$-th period, and $\eta_{ijk}$ is the linear predictor. For a continuous outcome,   an identity link function is often used for $g(\cdot)$ and the model is referred to as a linear mixed model. For a binary outcome, the logit link function is most often used, and the model is referred to as a logistic generalized linear mixed model.

\subsection{Immediate treatment effect models}
The standard GLMM for SW-CRT, which assumes a time-invariant immediate treatment effect, specifies the linear predictor as:
\begin{align}
\eta_{ijk} &= \mu + \beta_j + \delta X_{ij} + \mathcal{R}_{ij}, \label{eq:main-model}
\end{align}
\noindent where $X_{ij}$ is a binary treatment indicator for the $i$-th cluster in the $j$-th period ($X_{ij} = 1$ if the cluster period is under the intervention condition; $X_{ij} = 0$ if the cluster period is under the control condition). The overall mean under the control condition is denoted as $\mu$, and $\beta_j$ denotes a categorical fixed time effect ($j = 1, \dots, J$), with $\beta_1=0$ for identifiability, controlling for the secular trend. The parameter of interest (treatment effect) is denoted as $\delta$ and represents a constant treatment effect that does not vary over time. 

The term $\mathcal{R}_{ij}$ is a set of cluster-level and/or cluster-period level random effects to account for the dependence of observations within a cluster \cite{li_mixed-effects_2021}. For a continuous outcome (LMM), the model for the observation is $Y_{ijk} = \eta_{ijk} + \epsilon_{ijk}$, where $\epsilon_{ijk} \sim \mathcal{N}(0, \sigma^2)$ is the individual-level residual error. For a binary outcome (GLMM), the linear predictor $\eta_{ijk}$ is linked to the probability $p_{ijk}$ via the logit link, and the individual-level variance is determined by the Bernoulli distribution.

There are three common types of structures that $\mathcal{R}_{ij}$ can take under a cross-sectional stepped-wedge design. The EXCH correlation structure assumes a single cluster-level random effect. That is, $\mathcal{R}_{ij} = u_i$, and $u_i \sim \mathcal{N}(0, \sigma_u^2)$. 
This implies a constant correlation for all observations within the same cluster across multiple periods. For a continuous outcome, the ICC can be defined as
\begin{align}
ICC &= \frac{\sigma_u^2}{\sigma_u^2 + \sigma^2}  \nonumber
\end{align}
For a binary outcome, the ICC can be expressed on the logistic scale by replacing $\sigma^2$ with $\pi^2/3$. Please also see Ouyang et al. for additional details.\cite{ouyang_estimating_2023}

While simple, the EXCH correlation structure may not adequately reflect the longitudinal nature of SW designs. To introduce more flexibility, the NE structure incorporates an additional (independent) cluster-by-period random intercept \cite{hooper_sample_2016, girling_statistical_2016}, given by $\mathcal{R}_{ij} = u_i + v_{ij}$, $u_i \sim \mathcal{N}(0, \sigma_u^2)$ and $v_{ij} \sim \mathcal{N}(0, \sigma_v^2)$. 
This structure allows the model to differentiate between the within-period intracluster correlation (WP-ICC) and the between-period intracluster correlation (BP-ICC). This distinction is important because measurements from the same cluster and period can be more correlated than measurements from the same cluster but from different periods. In linear mixed models, the WP-ICC and BP-ICC can be expressed as (for continuous outcomes):
\begin{align*}
\text{WP-ICC} = \frac{\sigma_{u}^2 + \sigma_{v}^2}{\sigma_{u}^2 + \sigma_{v}^2 + \sigma^2},~~~~~
\text{BP-ICC} = \frac{\sigma_{u}^2}{\sigma_{u}^2 + \sigma_{v}^2 + \sigma^2},
\end{align*}
where $\sigma_{u}^2$ is the variance of the cluster-level random effect, $\sigma_{v}^2$ is the variance of the cluster-period random effect, and $\sigma^2$ is the individual-level residual variance (replace $\sigma^2$ by $\pi^2/3$ for binary outcomes).\cite{ouyang_estimating_2023} The relationship between these two correlations is quantified by the cluster autocorrelation coefficient (CAC), defined as:
\begin{equation}
CAC = \rho = \frac{\text{BP-ICC}}{\text{WP-ICC}} \nonumber \label{eq:cac}
\end{equation}
As the CAC is a constant under the NE structure, the correlation decreases from the WP-ICC to the BP-ICC for any two observations in different periods, and does not decay further as the temporal separation increases \cite{hooper_sample_2016}.

Next, the exponential decay (ED) structure assumes that the BP-ICC decays further with each additional measurement period apart \cite{kasza_impact_2019}. Specifically, the correlation structure is parameterized as $\mathcal{R}_{ij} = \gamma_{ij}$, $\boldsymbol{\gamma}_{i} = (\gamma_{i1}, \ldots, \gamma_{iJ}) \overset{\text{i.i.d.}}{\sim} \mathcal{N}_{J}(0, \sigma_{\gamma}^2 \mathcal{Z})$. 
Here, $\gamma_{ij}$ represents the cluster-period random effect for cluster $i$ in period $j$, with a common variance $\sigma_{\gamma}^{2}$ and covariance matrix $\sigma_{\gamma}^{2}\mathcal{Z}$. In general, the matrix $\mathcal{Z}$ is specified as an autoregressive order-1 (AR(1)) correlation structure:
\[
\mathcal{Z} = \mathcal{Z}(1, \rho) =
\begin{bmatrix}
1 & \rho & \rho^2 & \cdots & \rho^{J-1} \\
\rho & 1 & \rho & \cdots & \rho^{J-2} \\
\vdots & \vdots & \vdots & \ddots & \vdots \\
\rho^{J-1} & \rho^{J-2} & \rho^{J-3} & \cdots & 1 \\
\end{bmatrix}
\]
This specification introduces time-indexed cluster-period random effects, where the BP-ICC decays exponentially with increasing temporal separation between periods at a rate determined by $\rho$. Such modeling provides a more realistic representation of longitudinal dependence commonly observed in SW-CRT designs. The WP-ICC and BP-ICC can be expressed as:
\begin{align*}
\text{WP-ICC} = \frac{\sigma_\gamma^2}{\sigma_\gamma^2 + \sigma^2}, ~~~~~
\text{BP-ICC} = \frac{\sigma_\gamma^2 \cdot \rho^{|j - j'|}}{ \sigma_\gamma^2 + \sigma^2}.
\end{align*}

Finally, to account for treatment effect heterogeneity across clusters, any of the preceding correlation structures can be augmented with a random intervention effect \cite{voldal_model_2022}. This is typically specified as a cluster-level random effect, $t_i \sim \mathcal{N}(0, \sigma_t^2)$, which interacts with the treatment indicator. When included, $t_i$ may be assumed to be correlated with $u_i$. As an example, a Nested Exchangeable model with a Random Intervention (NE-RI) can be specified as:
\begin{align}
\mathcal{R}_{ij} &= u_i + v_{ij} + t_i X_{ij}.
\end{align}
This $t_i$ component, with variance $\sigma_t^2$, allows the intervention effect to differ across clusters and is used in the data-generating models in our simulation study to investigate the impact of correlation model specifications.

\subsection{Exposure-time indicator models}
In practice, it is possible that the treatment effect varies according to the exposure time (the number of periods exposed under the intervention).\cite{hughes_current_2015} Let $I_{ij}^{(e)}=1$ if the cluster \(i\) at time \(j\) has been exposed to the intervention for exactly \(e\) periods, and 0 otherwise. The ETI model considers a specific treatment effect parameter for each level of exposure time. Model (2) is specified as \cite{hughes_current_2015,kenny_analysis_2022,maleyeff_assessing_2023}:
\begin{equation}
\eta_{ijk} = \mu + \beta_j + \sum_{e=1}^{J-1} \delta_e\,I_{ij}^{(e)} + \mathcal{R}_{ij}.
\end{equation}
\noindent Here, $\delta_e$ is defined as the treatment effect after $e$ periods of exposure, with $e = (1,..., J-1)$ in a standard design with one full-control period and one full-intervention period. The other parameters are interpreted the same way as in the IT model. Under this model, the time-averaged treatment effect (TATE) is used as a single summary of the treatment effect and can be written as
\begin{equation}
\text{TATE} = \frac{1}{J-1}\sum_{e=1}^{J-1}\,\delta_e \nonumber
\end{equation}
Apart from this single summary, the ETI model also allows us to make inferences for any duration of treatment exposure. For example, investigators may be interested in the long-term treatment effect (LTE), defined as the effect after the longest duration of exposure $\delta_{J-1}$. Just as with Model (1), Model (2) specifies the linear predictor within the GLMM framework and can be used for both continuous and binary outcomes. We refer to Kenny et al.\cite{kenny_analysis_2022} and Wang et al.\cite{wang_anticipation_2026} for additional details and theoretical results of the IT models when the true data-generating model is the ETI model (misspecification of the treatment effect structure in the terminology of Wang et al.\cite{wang_how_2024}). 

\section{Cluster-Robust Variance Estimators}
The choice of random-effects structure is a common challenge in practice. For both LMM and GLMM, a misspecified structure can lead to inflated Type I error rates \cite{ouyang_sample_2022,ouyang_accounting_2023,ouyang_estimating_2023}. Under potential random-effects misspecification, RVEs provide a method for obtaining asymptotically valid SEs without requiring a correctly specified within-cluster covariance structure \cite{liang_longitudinal_1986}. By separating the estimation of regression parameters from the specification of the variance, RVEs enhance the reliability of inference in settings with correlated data.

For LMM with continuous outcomes, several small-sample corrected RVEs are readily available in the \texttt{clubSandwich} R package \cite{pustejovsky_clubsandwich_2022}. However, implementations for GLMM (e.g., for non-Gaussian outcomes) are not currently available in a publicly released R package. Therefore, to evaluate RVEs in the non-Gaussian setting considered here, we implemented the required estimators by writing custom functions that reproduce the corresponding sandwich forms after the GLMM linearization adapted from the SAS GLIMMIX procedure. The code is available on GitHub (https://github.com/jphughes9/vcovCRglmer).

In this study, we focus on estimators that have shown promise in previous research \cite{ouyang_maintaining_2024}. To formally introduce these estimators, we write the GLMM for each cluster $i$ (where $i=1, \dots, I$) as:

\begin{equation}
g\!\left(E[\mathbf{y}_i \mid \mathbf{b}_i]\right) = \mathbf{X}_i \boldsymbol{\beta} + \mathbf{Z}_i \mathbf{b}_i,
\end{equation}
where $\mathbf{y}_i$ is the vector of outcomes for the $k=1,\dots,K_i$ individuals across $j=1,\dots,J$ periods in cluster $i$. $\mathbf{X}_i$ and $\mathbf{Z}_i$ are the design matrices for the fixed and random effects, respectively, $\boldsymbol{\beta}$ is the vector of fixed effects, and $\mathbf{b}_i$ is the vector of random effects. The random effects are typically assumed to be drawn from a normal distribution $\mathbf{b}_i \sim \mathcal{N}(\mathbf{0}, \mathbf{R}(\boldsymbol{\theta}))$.

\paragraph{Linearization for non-Gaussian outcomes}
Let $\boldsymbol{\eta}_i = \mathbf{X}_i\boldsymbol{\beta} + \mathbf{Z}_i\mathbf{b}_i$ and $\boldsymbol{\mu}_i = g^{-1}(\boldsymbol{\eta}_i)$. Let $\boldsymbol{\Sigma}_i$ denote the conditional variance of $\mathbf{y}_i$ given $\mathbf{b}_i$ (e.g., diagonal with elements determined by the mean--variance relationship. For binomial outcomes entered in aggregated form, $\boldsymbol{\Sigma}_i$ must incorporate the number of trials). Define the diagonal matrix of derivatives
\[
\boldsymbol{\Delta}_i = \left.\frac{\partial g^{-1}(\boldsymbol{\eta}_i)}{\partial \boldsymbol{\eta}_i}\right|_{(\boldsymbol{\beta},\mathbf{b}_i)}.
\]
The working (pseudo-)response is
$
\mathbf{P}_i
=
\boldsymbol{\Delta}_i^{-1}\left(\mathbf{y}_i - \boldsymbol{\mu}_i\right) + \boldsymbol{\eta}_i,
$
and the corresponding working marginal covariance matrix is
\begin{equation}
\mathbf{V}_i
=
\mathbf{Z}_i\,\mathbf{R}(\boldsymbol{\theta})\,\mathbf{Z}_i^\top
+
\boldsymbol{\Delta}_i^{-1}\boldsymbol{\Sigma}_i\,\boldsymbol{\Delta}_i^{-1}.
\end{equation}
All quantities are evaluated at the fitted values $(\hat{\boldsymbol{\beta}},\hat{\boldsymbol{\theta}},\hat{\mathbf{b}}_i)$, yielding $\widehat{\mathbf{V}}_i$ and $\hat{\mathbf{P}}_i$. The linearized residual vector used in the sandwich ``meat'' is
$
\mathbf{e}_i = \hat{\mathbf{P}}_i - \mathbf{X}_i \hat{\boldsymbol{\beta}}.
$
This construction reduces to the standard LMM when $g$ is the identity link (so $\boldsymbol{\Delta}_i=\mathbf{I}$ and $\mathbf{P}_i=\mathbf{y}_i$).

\paragraph{Classic sandwich estimator}
The standard sandwich estimator ($\widehat{\mathbf{V}}_{\mathrm{classic}}$) provides consistent variance estimates in large samples but can be biased downwards when the number of clusters is small, potentially inflating Type I error rates \cite{zeger_longitudinal_1986}. It applies no finite-sample correction.
Let
$
\mathbf{M} = \sum_{i=1}^I \mathbf{X}_i^\top \widehat{\mathbf{V}}_i^{-1}\mathbf{X}_i,
$ 
the classic sandwich estimator can be written as: 

\begin{equation}
\widehat{\mathbf{V}}_{\mathrm{classic}}
=
\mathbf{M}^{-1}
\left(
\sum_{i=1}^I
\mathbf{X}_i^\top \widehat{\mathbf{V}}_i^{-1}\,
\mathbf{e}_i \mathbf{e}_i^\top\,
\widehat{\mathbf{V}}_i^{-1} \mathbf{X}_i
\right)
\mathbf{M}^{-1}.
\end{equation}

\paragraph{Kauermann and Carroll (Bias-corrected estimator)}
The Kauermann and Carroll (KC) estimator, originally proposed by Bell and McCaffrey \cite{bell_bias_2002}, adjusts the residuals using a leverage-like matrix to account for uncertainty in the estimated mean parameters. This correction reduces the small-sample bias of $\hat{V}_{classic}$ and is generally recommended when the number of clusters is moderate (e.g., fewer than 50). For this estimator, we define

\[
\mathbf{H}_i
=
\mathbf{X}_i\,\mathbf{M}^{-1}\mathbf{X}_i^\top\,\widehat{\mathbf{V}}_i^{-1}
\qquad (n_i \times n_i),
\]
and the KC residual adjustment matrix is:
\[
\mathbf{F}_{i,\mathrm{KC}} = \left(\mathbf{I}_{n_i} - \mathbf{H}_i^\top\right)^{-1/2},
\]
where $(\cdot)^{-1/2}$ denotes the inverse of a matrix square root. The KC estimator is then
\begin{equation}
\widehat{\mathbf{V}}_{\mathrm{KC}}
=
\mathbf{M}^{-1}
\left(
\sum_{i=1}^I
\mathbf{X}_i^\top \widehat{\mathbf{V}}_i^{-1}\,
\mathbf{F}_{i,\mathrm{KC}}\mathbf{e}_i \mathbf{e}_i^\top \mathbf{F}_{i,\mathrm{KC}}^\top\,
\widehat{\mathbf{V}}_i^{-1} \mathbf{X}_i
\right)
\mathbf{M}^{-1}.
\end{equation}

\paragraph{Mancl and DeRouen (approximate jackknife estimator)}
The Mancl and DeRouen (MD) estimator is analogous to a leave-one-cluster-out jackknife procedure, which provides a more conservative variance estimate. By capturing the influence of each cluster on the overall parameter estimates, it offers stronger Type I error control in very small-sample contexts. This estimator is equivalent to that proposed by Mancl and DeRouen in the GEE literature \cite{mancl_covariance_2001}. Using the same $\mathbf{H}_i$ as above, define: 
\[
\mathbf{F}_{i,\mathrm{MD}} = \left(\mathbf{I}_{n_i} - \mathbf{H}_i^\top\right)^{-1}.
\]
The MD estimator is
\begin{equation}
\widehat{\mathbf{V}}_{\mathrm{MD}}
=
\mathbf{M}^{-1}
\left(
\sum_{i=1}^I
\mathbf{X}_i^\top \widehat{\mathbf{V}}_i^{-1}\,
\mathbf{F}_{i,\mathrm{MD}}\mathbf{e}_i \mathbf{e}_i^\top \mathbf{F}_{i,\mathrm{MD}}^\top\,
\widehat{\mathbf{V}}_i^{-1} \mathbf{X}_i
\right)
\mathbf{M}^{-1}.
\end{equation}

\paragraph{Morel, Bokossa, and Neerchal estimator}
We also evaluated one additional bias-corrected RVE that is not included in our main \texttt{clubSandwich}-based workflow. Morel, Bokossa, and Neerchal \cite{morel_small_2003} proposed an alternative small-sample correction, primarily for the GEE context. The Morel, Bokossa, and Neerchal (MBN) estimator adjusts the standard sandwich estimator ($\widehat{\mathbf{V}}_{\mathrm{classic}}$) with both a leading scaling factor and an additive inflation term, yielding $\widehat{\mathbf{V}}_{\mathrm{MBN}}$. Let $N$ be the total number of observations and $p$ the number of fixed-effect parameters. The MBN estimator is

\begin{equation}
\widehat{\mathbf{V}}_{\mathrm{MBN}}
=
c \cdot \widehat{\mathbf{V}}_{\mathrm{classic}}
+
\delta_I \,\phi\; \mathbf{M}^{-1},
\end{equation}
with scaling factor
$
c = \frac{N-1}{N-p}\cdot\frac{I}{I-1}.
$
To define the additive term, let
\begin{equation}
\boldsymbol{\Omega}
=
\mathbf{M}^{-1}
\left(
\sum_{i=1}^I
\mathbf{X}_i^\top \widehat{\mathbf{V}}_i^{-1}\,
\mathbf{e}_i\mathbf{e}_i^\top\,
\widehat{\mathbf{V}}_i^{-1}\mathbf{X}_i
\right),
\end{equation}
and define $p^\ast=p$ if $I>p$, otherwise let $p^\ast$ be the number of nonzero singular values of $\boldsymbol{\Omega}$. Then
$
\phi = \max\!\left(r_{\mathrm{MBN}},\,\frac{\mathrm{tr}(\boldsymbol{\Omega})}{p^\ast}\right),
$
and
\[
\delta_I=
\begin{cases}
\dfrac{p}{I-p}, & \text{if } I>(d+1)p,\\[6pt]
\dfrac{1}{d}, & \text{otherwise},
\end{cases}
\]
where $d\ge 1$ (with default $d = 2$) and $0\le r_{\mathrm{MBN}}\le 1$ are user-specified. The additive component vanishes as the number of clusters increases, but can improve small-sample performance when $I$ is small relative to $p$.

Empirical studies have repeatedly shown that RVEs alone are insufficient to ensure valid inference when the number of clusters is small \cite{ford_maintaining_2020,fay_small-sample_2001,pustejovsky_small-sample_2018}. Therefore, their use is typically combined with use of a $t$-distribution with a small-sample degrees of freedom correction for statistical inference.

\section{Simulation study}
This section details a comprehensive simulation study designed to empirically evaluate the performance of RVEs for the ETI model. Our primary objectives are to determine whether RVEs, including various small-sample corrections, provide valid inference for time-varying treatment effects, particularly when the random-effects structure is misspecified. We pay special attention to designs with a limited number of clusters, aiming to identify the most reliable methods for both continuous and binary outcomes.

\subsection{Simulation design}

For continuous outcomes, this study builds upon previous research of Ouyang et al. demonstrating the utility of RVEs for IT models \cite{ouyang_maintaining_2024}. We extend this to the ETI model by generating data from a highly complex temporal structure, the Exponential Decay with Random Intervention (ED-RI), which is computationally feasible within the linear mixed-effects model framework. We then assess the robustness of RVE when simpler, misspecified models are fitted, as such simpler random-effects models are frequently used in current practice.\cite{nevins_adherence_2024}

Conversely, the empirical performance of RVE for binary outcomes is less explored. Fitting GLMMs with complex random-effects structures like ED-RI is often computationally intensive and prone to instability. We therefore adopt a more foundational approach. For binary outcomes, this foundational approach should be interpreted as a benchmark evaluation of RVE behavior under ETI-based model fitting, rather than as a full assessment. We generate data from established structures, including the EXCH, NE, and NE with random intervention (NE-RI) models. Our considerations are twofold. First, we evaluate performance under ideal conditions by fitting the correctly specified model and assessing the performance characteristics of the RVE estimator. Second, we investigate robustness by fitting a misspecified EXCH model to data generated from these more complex structures. We would like to acknowledge that a potential caveat that the non-collapsibility of effect measures for binary outcomes may induce discrepancies between estimands in different GLMMs even under correct fixed-effects specification. Nevertheless, our goal is not to eliminate this issue by design, but rather to empirically assess the extent to which such non-collapsibility manifests in finite samples and how it ultimately propagates to the RVE-based inference in SW-CRTs. This perspective allows us to evaluate the practical reliability of RVE in realistic settings when binary outcomes are considered. 

Across all scenarios, the study assesses the validity of inference for TATE and LTE, focusing on identifying the optimal RVE for designs with a limited number of clusters. All models were fitted in R using \texttt{lmer} via \texttt{lmerTest} package for continuous outcomes and \texttt{glmer} via \texttt{lme4} package for binary outcomes.

\subsection{Data-generating process}
We considered a standard cross-sectional SW-CRT design with $I$ clusters, $J$ periods, and $K$ individuals per cluster-period. Key design parameters were varied to create a range of realistic scenarios \cite{nevins_adherence_2024}.

\subsubsection{Parameters for continuous outcomes}
The simulation parameters for the continuous outcome scenario are detailed in Table~\ref{tab:combined_params}. The number of clusters ($I$) ranged from 8 to 32, the number of periods ($J$) was set as either 5 or 9, while cluster-period sizes ($K$) were set as 10 or 50. The true treatment effect ($\delta_e$) was modeled as an equally spaced increasing linear trend, starting from a reference effect of 0 (no exposure) and increasing to 1 at the final exposure period ($e=J-1$). For example, when $J=5$, the set of four estimated parameters is $(\delta_1, \delta_2, \delta_3, \delta_4) = (0.25, 0.50, 0.75, 1.0)$. The underlying time trend was set at zero (i.e. $\beta_j = 0$ for all $j$)

Furthermore, we specified three scenarios for the random effect standard deviations based on an ED-RI structure (following Ouyang et al. \cite{ouyang_maintaining_2024}). As shown in the table, these scenarios vary the cluster-period SD ($\sigma_\gamma$) and the random intervention SD ($\sigma_t$) while holding the decay parameter (autocorrelation: $\rho$) constant at 0.8. This creates different WP-ICCs in the control and intervention groups. The residual error standard deviation ($\sigma$) was fixed at 1.

\begin{table}[!htbp]
\caption{Simulation scenarios for continuous and binary outcomes}
\label{tab:combined_params}
\centering
\renewcommand{\arraystretch}{1.12}
\setlength{\tabcolsep}{5pt}

\begin{threeparttable}
\begin{tabular}{@{} l l l @{}}
\toprule

\multicolumn{3}{@{}l}{\textbf{General design parameters}}\\
\midrule
Clusters ($I$) & Periods ($J$) & Cluster-period size ($K$)\\
\addlinespace[2pt]
$8,16,32$ & $5,9$ & $10,50$\\

\midrule
\multicolumn{3}{@{}l}{\textbf{Continuous outcome}}\\
\midrule
Effect-size trend & \multicolumn{2}{l}{Linear sequence $0 \to 1$}\\
Residual SD ($\sigma$) & \multicolumn{2}{l}{1}\\
Autocorrelation ($\rho$) & \multicolumn{2}{l}{0.8}\\
\addlinespace[4pt]
\multicolumn{3}{@{}l}{\textit{Random-effects settings}}\\
\addlinespace[2pt]
Scenario (ED-RI) & \multicolumn{2}{l}{SDs $(\sigma_\gamma,\sigma_t)$}\\
\addlinespace[2pt]
I   & \multicolumn{2}{l}{$(0.10, 0.21)$}\\
II  & \multicolumn{2}{l}{$(0.23, 0.24)$}\\
III & \multicolumn{2}{l}{$(0.23, 0.35)$}\\

\midrule
\multicolumn{3}{@{}l}{\textbf{Binary outcome}}\\
\midrule
Baseline probability ($p_0$) & \multicolumn{2}{l}{0.2, 0.5}\\
Treatment effect (log-OR) & \multicolumn{2}{l}{0.25}\\
Time effect (log-OR) & \multicolumn{2}{l}{0}\\
\addlinespace[4pt]
\multicolumn{3}{@{}l}{\textit{Random-effects settings}}\\
\addlinespace[2pt]
Scenario & \multicolumn{2}{l}{SDs $(\sigma_u,\sigma_v,\sigma_t)$}\\
\addlinespace[2pt]
I  (EXCH)   & \multicolumn{2}{l}{$(0.42, 0, 0)$}\\
II (NE)     & \multicolumn{2}{l}{$(0.37, 0.19, 0)$}\\
III (NE-RI) & \multicolumn{2}{l}{$(0.37, 0.19, 0.2)$}\\

\bottomrule
\end{tabular}

\begin{tablenotes}[flushleft]
\footnotesize
\item EXCH = exchangeable; NE = nested exchangeable; RI = random intervention effect.
\end{tablenotes}
\end{threeparttable}
\end{table}

\subsubsection{Parameters for binary outcomes}
Simulation parameters for the binary settings are detailed in Table~\ref{tab:combined_params}. The design parameters ($I, J, K$) were varied identically to the continuous settings. We varied the baseline event probability (0.2 and 0.5). The true treatment effect was a constant log-odds ratio (OR) of 0.25 for all exposure periods, with secular time trend log-OR = 0. For binary outcomes, we did not generate data under genuine exposure-time treatment effects. Instead, we used a constant log-OR of 0.25 across exposure periods as a calibration setting to isolate the finite-sample behavior of the variance estimators under ETI-based mixed-model fitting. Introducing a truly time-varying effects in logistic mixed models would make empirical assessment of bias and coverage considerably harder to interpret, because poor performance could then arise from a combination of sparse exposure-specific cells, instability in estimating late exposure effects, and the non-collapsibility of OR-based estimands across mixed-model specifications. By holding the treatment effect constant and setting the secular time trend to zero, we intentionally simplified the mean structure so that differences across methods are driven primarily by the random-effects correlation structure and the small-sample behavior of the variance estimators. As shown in Table~\ref{tab:combined_params}, the three random effect scenarios were chosen to represent an EXCH, a NE, and a NE-RI correlation structure.

\subsection{Performance measures}
We assessed model performance across 2,000 simulations per scenario. Our primary measure was the empirical coverage probability of the 95\% confidence intervals (CIs) for the TATE and LTE, which is defined as the proportion of simulations runs where the true effect ($\delta$) falls between the lower ($\hat{\delta}_{\text{low}}$) and upper ($\hat{\delta}_{\text{upp}}$) bounds of the 95\% CI of the estimated treatment effects (e.g., $P(\hat{\delta}_{\text{low}} \le \delta \le \hat{\delta}_{\text{upp}})$). We also calculated the bias of these estimates. CIs were constructed using both model-based and all RVEs, with inference based on both a Normal distribution and a $t$-distribution with $I-2$ degrees of freedom.


\section{Results}

\subsection{Continuous outcomes}
The results presented derive from fitting a misspecified EXCH correlation structure to data generated from an ED-RI structure.

\subsubsection{Performance for the time-averaged treatment effect}
\begin{sidewaysfigure}
    \centering
    \begin{subfigure}{0.48\textheight}
        \centering
        \includegraphics[width=\linewidth]{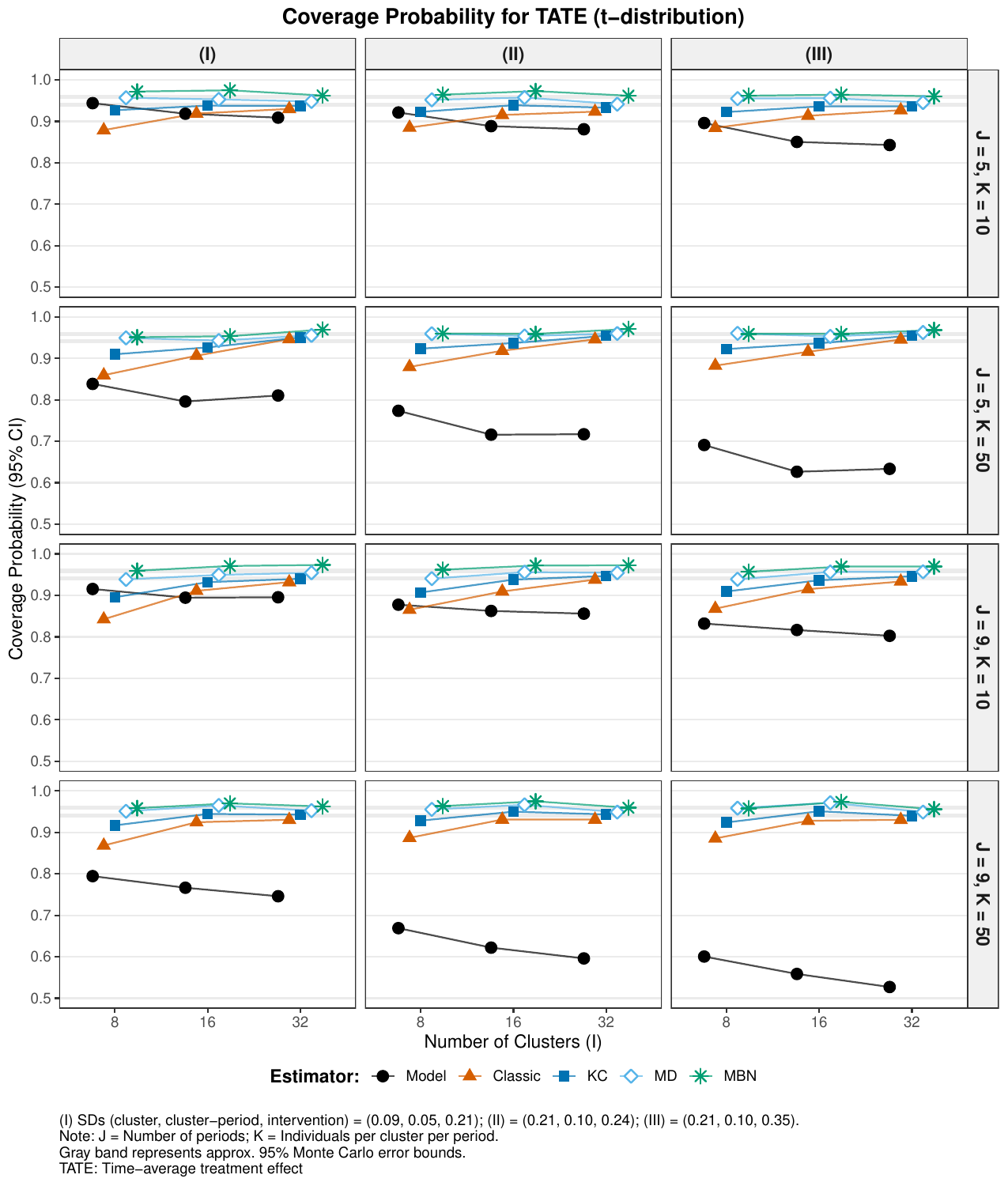}
        \caption{Inference using a $t$-distribution}
        \label{fig:cont_tate_t}
    \end{subfigure}\hfill
    \begin{subfigure}{0.48\textheight}
        \centering
        \includegraphics[width=\linewidth]{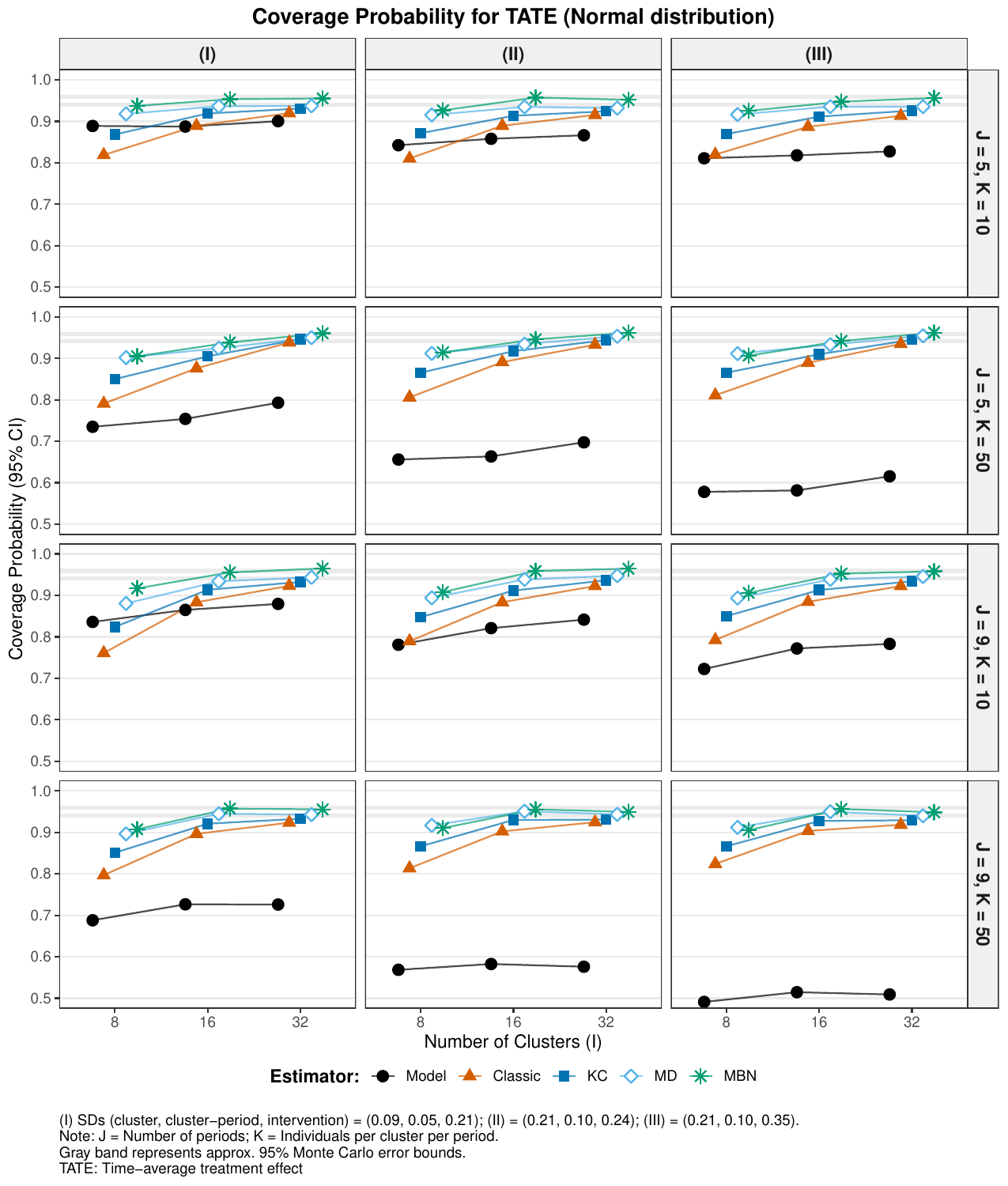}
        \caption{Inference using a Normal approximation}
        \label{fig:cont_tate_z}
    \end{subfigure}

    \caption{Coverage probability of nominal 95\% confidence intervals for the time-averaged treatment effect (TATE) in continuous outcome simulations under misspecified random-effects structures. The left panel (a) uses $t$-distribution critical values, and the right panel (b) uses Normal-approximation critical values. Within each subplot, coverage is plotted against the number of clusters $I\in\{8,16,32\}$ for five variance estimators (Model, Classic, KC, MD, MBN). Rows correspond to the number of periods $J$ and individuals per cluster-period $K$ (top to bottom: $(J,K)=(5,10),(5,50),(9,10),(9,50)$). Columns (I)--(III) correspond to variance-component settings $(\sigma_\gamma,\sigma_t) = (0.10,0.21), (0.23,0.24), (0.23,0.35)$. The dashed horizontal line marks the 0.95 nominal level, and the shaded band indicates approximate 95\% Monte Carlo error bounds.}
    \label{fig:cov_continuous}
\end{sidewaysfigure}

While TATE estimates were virtually unbiased across all scenarios (See Web Appendix A, Web Table 1), CI coverage was highly sensitive to the choice of SE estimator (Figure~\ref{fig:cov_continuous}). Model-based CIs performed poorly under misspecification, with coverage deteriorating well below 80\% for larger trials.

CIs based on RVEs demonstrated more reliable performance overall, though notable differences were observed among different RVEs. Despite this improvement, the classic RVE estimator consistently showed under-coverage, even when the number of clusters was 32. Among the small-sample correction estimators, the MD paired with a $t$-distribution provided the most accurate and stable coverage across nearly all scenarios (Figure~\ref{fig:cont_tate_t}). The coverage closely matched the nominal 95\% level, even in smaller trials, and showed minimal sensitivity to the number of clusters or cluster-period size.  Performance depended on the choice of distribution for calculating  CIs. The MD estimator, combined with a Normal approximation (Figure~\ref{fig:cont_tate_z}, was markedly anti-conservative (with coverage below 90\%), particularly in the case of a small number of clusters (e.g., $I=8, J=9$). However, the coverage improved with 32 clusters.

The MBN estimator also showed strong performance. When used with a $t$-distribution (Figure~\ref{fig:cont_tate_t}), it tended to be slightly conservative in larger trials (16 and 32 clusters), while the normal approximation corrected this conservatism, yielding coverage near the 95\% nominal level for $I \ge 16$. However, MBN (under normal approximation) displayed under-coverage in smaller designs (8 clusters) under the Normal approximation (as low as 90\%). Across all settings, both MD and MBN outperformed the classic and KC estimators.

\subsubsection{Performance for the long-term treatment effect}
Estimation of the LTE proved more challenging. A slight positive bias (See Web Appendix A, Web Table 1) was observed in scenarios with few clusters. Overall, the CI coverage for the LTE was poorer across all methods compared to the TATE. Under the t-distribution, MBN frequently provided conservative coverage, and MD provided coverage under 95\%, especially under the small number of clusters settings. Under the normal distribution, MD continues with undercoverage. MBN estimators were able to reach nominal coverage when $I \geq 16$, but failed to achieve nominal coverage when $I=8$ (Web Appendix A, Web Tables 2 and 3). Nevertheless, the general performance trends and relative rankings of the estimators were consistent with those observed for the TATE.

\subsection{Binary outcomes}
\subsubsection{Bias}
For binary outcomes, bias was small to negligible in most scenarios when $K=50$ or $I>8$. However, a challenging case emerged in designs with one-cluster-per-sequence (1CPS: $I=8$, $J=9$, $K=10$) combined with a low baseline event probability (e.g., 0.2), as illustrated in Web Appendix A, Web Table 4. In this specific design, the estimation of the LTE relies exclusively on data from the single cluster-period corresponding to the longest exposure duration. With a low baseline event probability and low cluster-period size, this critical cell has a high likelihood of observing zero events. This absence of events leads to unstable parameter estimation and a highly inflated LTE estimate. This extreme bias directly propagated to the TATE, which, as an average of all exposure-specific effects, exhibited a substantial bias. Because of the zero event, the LTE bias is effectively unbounded, so any finite bias observed in the simulations is a numerical artifact and its magnitude is not meaningful. This bias was substantially mitigated by increasing the baseline event probability to 0.5 (See Web Appendix A, Web Table 5), which ensured a sufficient number of events in the critical cells for stable estimation. Bias was also significantly reduced when we remove all non-convergence simulation (due to data sparsity) runs.

\subsubsection{Coverage probability for the time-averaged treatment effect}
\begin{sidewaysfigure}
    \centering
    \begin{subfigure}{0.48\textheight}
        \centering
        \includegraphics[width=\linewidth]{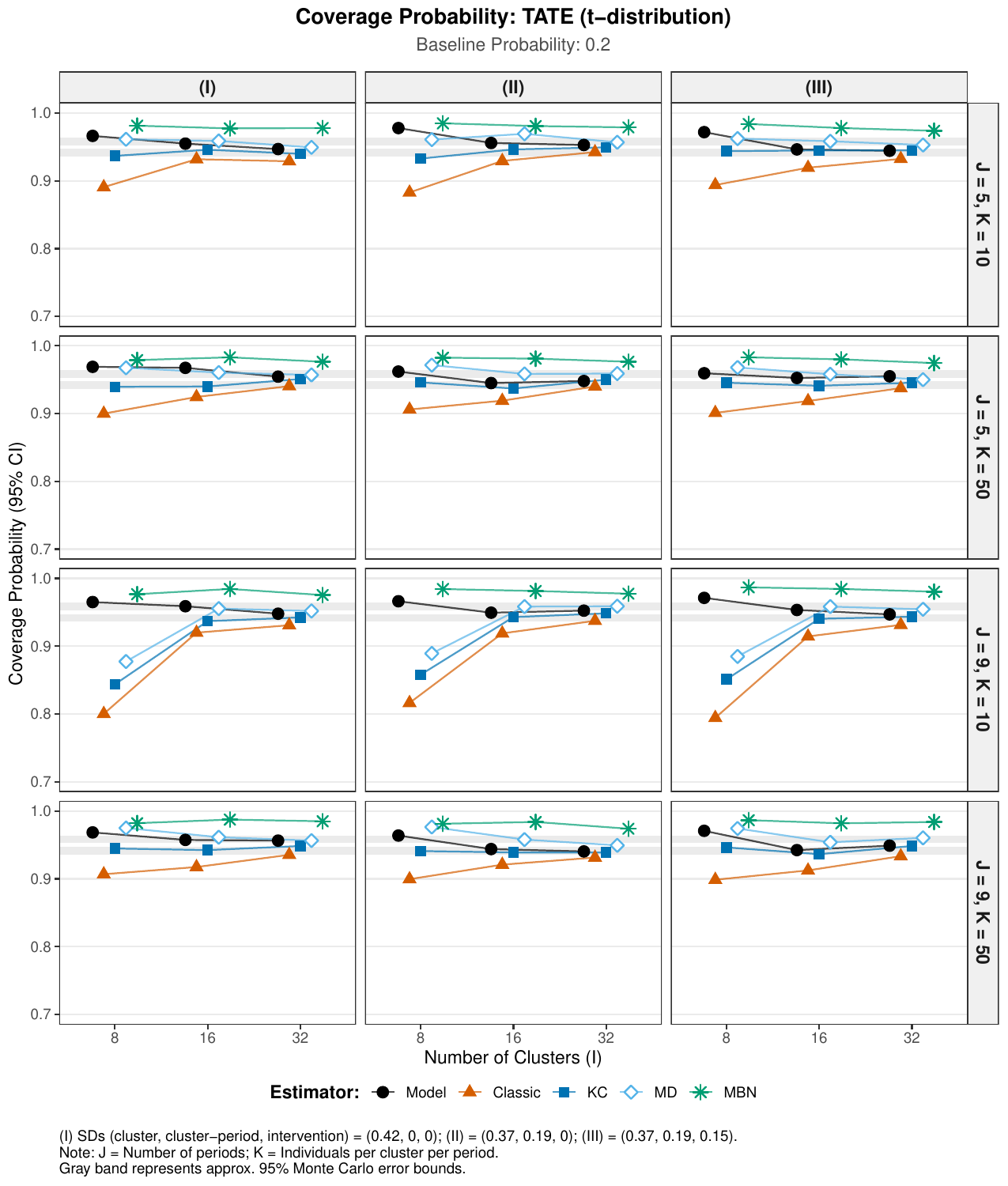}
        \caption{Inference using a $t$-distribution}
        \label{fig:cov_binary_0.2_t}
    \end{subfigure}\hfill
    \begin{subfigure}{0.48\textheight}
        \centering
        \includegraphics[width=\linewidth]{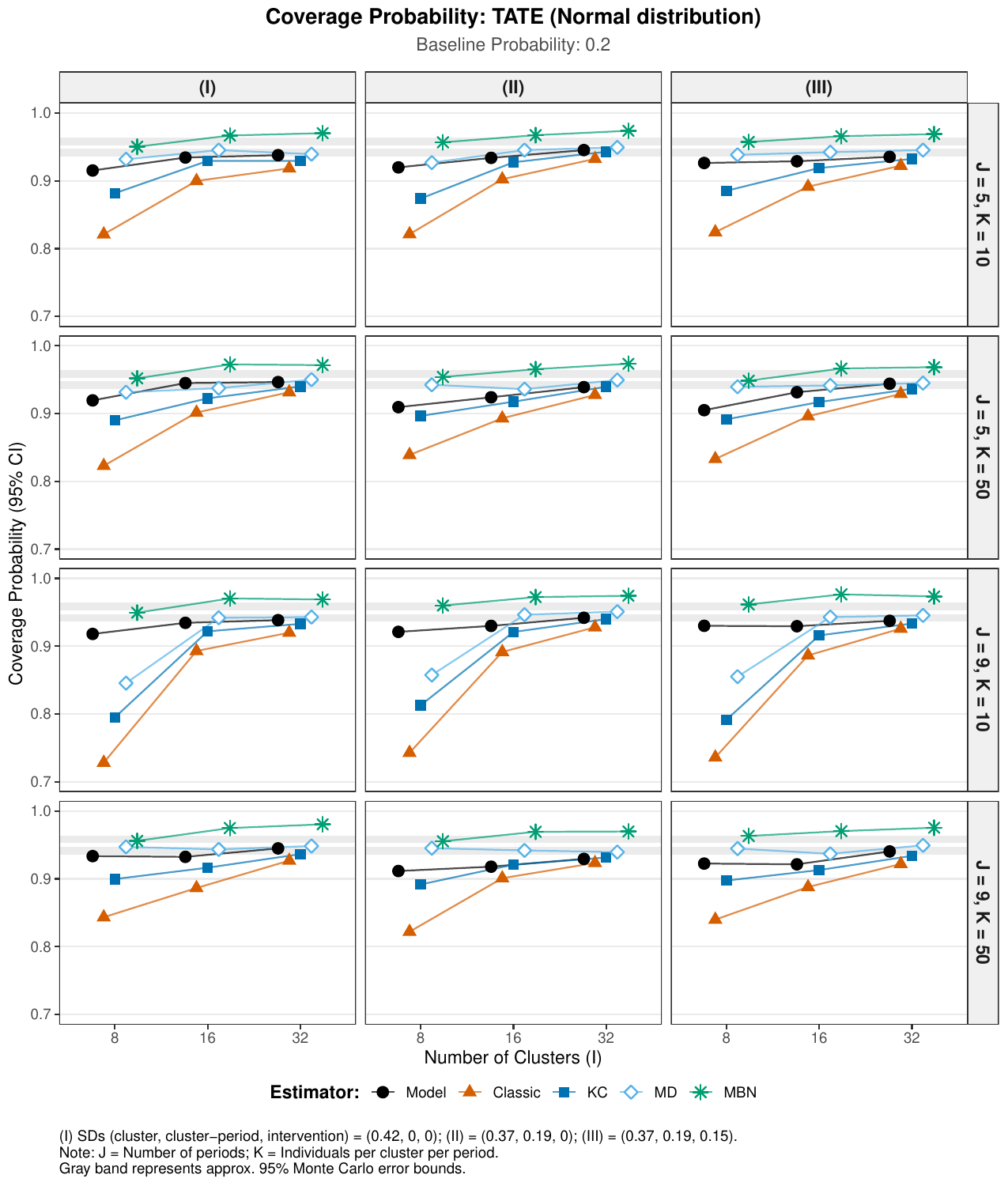}
        \caption{Inference using a Normal approximation}
        \label{fig:cov_binary_0.2_z}
    \end{subfigure}

    \caption{Coverage probability of nominal 95\% confidence intervals for the TATE in binary outcome simulations with baseline event probability $0.2$ under correctly specified correlation structures. The left panel (a) uses $t$-distribution critical values, whereas the right panel (b) uses Normal-approximation critical values. Within each subplot, coverage is shown as a function of the number of clusters $I\in\{8,16,32\}$ for five variance estimators (Model, Classic, KC, MD, MBN). Rows correspond to the number of periods $J$ and individuals per cluster-period $K$ (top to bottom: $(J,K)=(5,10),(5,50),(9,10),(9,50)$). Columns (I)--(III) correspond to variance-component settings for $(\sigma_u,\sigma_v,\sigma_t)=(0.42,0,0)$, $(0.37,0.19,0)$, and $(0.37,0.19,0.20)$, respectively. The dashed horizontal line marks the 0.95 nominal level and the shaded band indicates approximate 95\% Monte Carlo error bounds.}
    \label{fig:cov_binary}
\end{sidewaysfigure}

As with continuous outcomes, coverage probabilities depended on the choice of variance estimators and reference distributions. When the random-effects structure was correctly specified, the model-based SEs achieved nominal coverage rates (approximately 95\%) when utilizing a $t$-distribution (Figure~\ref{fig:cov_binary_0.2_t}). In these ideal scenarios, the model-based approach performed comparably to the small-sample corrected RVEs (e.g., KC and MD), which also maintained adequate coverage across cluster sizes, except for the scenario where we have $I=8, J=9$. The classic RVE suffered from undercoverage when the number of clusters was small, though its performance converged toward the 95\% nominal level as the number of clusters increased. However, when the standard normal approximation was used with a small number of clusters, only the MBN method achieved nominal coverage. All other methods resulted in undercoverage even under the correctly specified model. (Figure~\ref{fig:cov_binary_0.2_z}).

\begin{sidewaysfigure}
    \centering
    \begin{subfigure}{0.48\textheight}
        \centering
        \includegraphics[width=\linewidth]{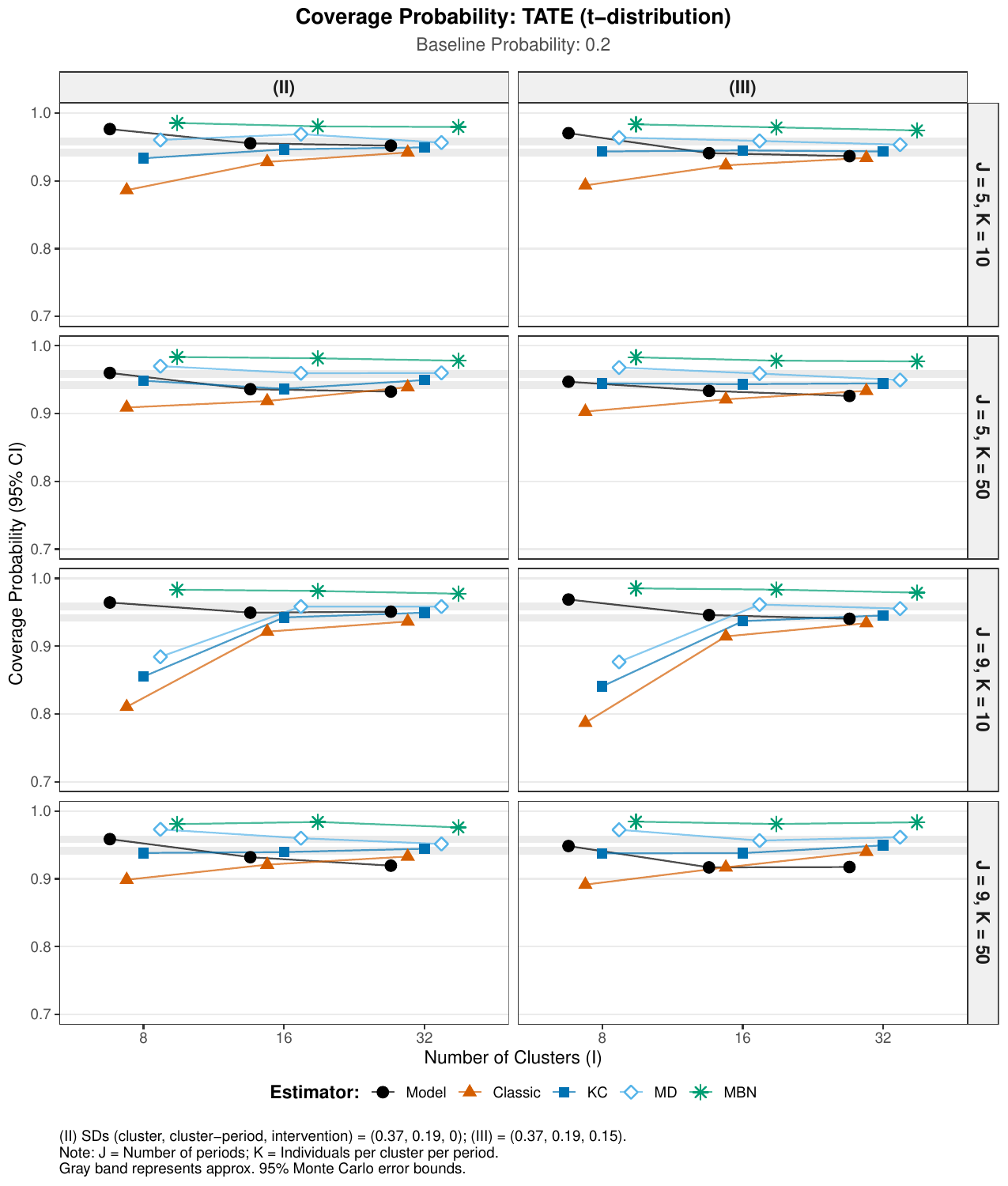}
        \caption{Inference using a $t$-distribution}
        \label{fig:cov_binary_0.2_t_mis}
    \end{subfigure}\hfill
    \begin{subfigure}{0.48\textheight}
        \centering
        \includegraphics[width=\linewidth]{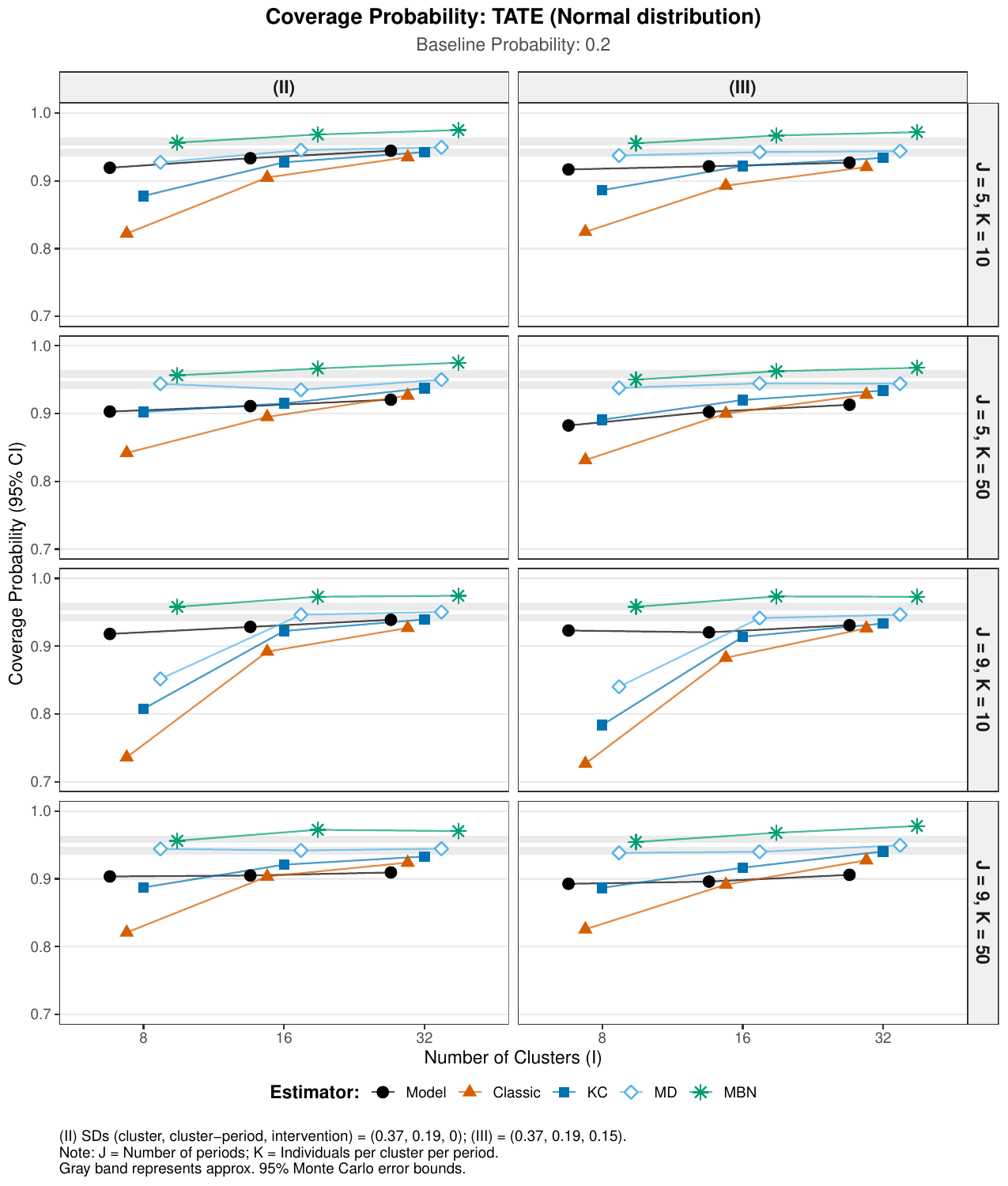}
        \caption{Inference using a Normal approximation}
        \label{fig:cov_binary_0.2_mis}
    \end{subfigure}

    \caption{Coverage probability of nominal 95\% confidence intervals for the time-averaged treatment effect (TATE) in binary-outcome simulations with baseline event probability $0.2$ when the fitted binary-outcome model is misspecified. The left panel uses $t$-distribution critical values, whereas the right panel uses Normal-approximation critical values. Within each subplot, coverage is plotted against the number of clusters $I\in\{8,16,32\}$ for five variance estimators (Model, Classic, KC, MD, MBN). Rows correspond to the number of periods $J$ and individuals per cluster-period $K$ (top to bottom: $(J,K)=(5,10),(5,50),(9,10),(9,50)$). Only variance-component settings (II)--(III) are shown: $(\sigma_u,\sigma_v,\sigma_t)=(0.37,0.19,0)$ and $(0.37,0.19,0.20)$. The dashed horizontal line marks the 0.95 nominal level, and the shaded band indicates approximate 95\% Monte Carlo error bounds.}
    \label{fig:cov_binary_mis}
\end{sidewaysfigure}

When the random-effects structure was misspecified, the coverage, in general, fell below the 95\% nominal level, whereas the MD and MBN estimators remained reliable, except for MD when $I=8$, $J=9$, and $K=10$. Notably, the performance of the RVE-based CIs was largely consistent across both correctly specified and misspecified models, underscoring their robustness to random-effects misspecification for most scenarios. The choice of reference distribution was also critical. When using a $t$-distribution (Figure~\ref{fig:cov_binary_0.2_mis}), the MBN estimator was consistently conservative (above 95\%), whereas the MD estimator generally provided nominal coverage but exhibited substantial under-coverage (dropping below 90\%) in challenging 1CPS designs. Using a Normal approximation (Figure~\ref{fig:cov_binary_0.2_mis}) induced a liberal shift in all estimators. This shift had a corrective effect on the MBN estimator, counteracting its conservatism and bringing its coverage closer to the nominal level. However, it caused the MD estimator to have undercoverage in most scenarios. Overall, the MBN estimator paired with a Normal approximation provided the best coverage probabilities, although it can be conservative in larger trials.

\subsubsection{Coverage probability for the long-term treatment effect}
Consistent with the continuous outcome results, CI performance for the LTE was uniformly poorer than for the TATE. Model-based CIs were highly unreliable under misspecification. The MD and MBN estimators, while superior, still struggled to achieve nominal coverage in scenarios with a small number of clusters (See Web Appendix A, Web Tables 6 and 7).

\subsubsection{A critical vulnerability: weak event support in the longest-exposure ETI cell}

Although earlier section suggested that poor performance was concentrated in 1CPS settings, our simulation results reveals a more general mechanism. Instability in ETI analyses appears to be driven less by the nominal sequence structure alone and more by the amount of outcome information supporting the longest treatment exposure cells. In this sense, 1CPS designs represent an extreme case, because the longest-exposure effect may be informed by very limited data, but the underlying vulnerability is broader. ETI models become fragile whenever the most longest exposed cell is weakly supported.

In Web Appendix A, Web Figure 1, we plotted the distribution of model-based, MD and MBN adjusted SEs over a few scenarios. In general, we found that the LTE SE is larger than the corresponding TATE SE. This difference is most pronounced when the number of events in the longest-exposure ETI cell is very small. The pattern is especially severe in the 8-cluster, 9-period settings, where information about exposure-time effects is spread thinly across many periods. In contrast, with 32 clusters and 5 periods, both TATE and LTE estimation are more stable, indicating that the problem is mitigated when support for the ETI structure is stronger.

The behavior of the RVEs also differs in these settings. The MD estimator shows substantial variability for both the TATE and LTE in settings with a small number of clusters settings, particularly under 1CPS settings. The greatest instability arises when the longest‑exposure cell contains only one or two events. Interestingly, when that cell contains zero events, the MD distributions return to a more typical range, whereas the model‑based SE and MBN distributions become so inflated that they fall outside the plotting range. However, this does not imply that the MD estimator should be used in these cases. We do not recommend making inferences that involve exposure time periods with zero events. Even when additional events are present, the MD estimator remains highly variable in small‑cluster settings. This pattern indicates that the estimator’s vulnerability is not driven solely by extreme event sparsity, but also by the sensitivity of jackknife‑type variance estimation to high‑leverage exposure indicators and limited between‑cluster information in small‑sample ETI models.

In contrast, the MBN estimator demonstrates substantially greater stability across event-count strata and design settings. Although LTE SEs remain larger than those for the TATE, the MBN estimator exhibits much less variability than MD and appears substantially less sensitive to the number of events in the longest-exposure cell. Collectively, these findings suggest that the practical warning for ETI analyses is not simply to avoid 1CPS designs, but more generally to assess whether the longest-exposure component of the model is likely to be supported by enough outcome events. When that support is weak, particularly in trials with few clusters and many periods, LTE estimation may be imprecise and MD-based inference may be unreliable. In such settings, investigators should be cautious about emphasizing long-exposure contrasts and should prefer more stable inferential approaches, such as MBN when binary outcomes are analyzed. If scientifically reasonable, one could reduce the number of ETI parameters by combining exposure times as suggested in Hughes et al. \cite{hughes_sample_2024}

\section{Illustrative Examples}
In this section, we illustrate how RVE can be implemented in a realistic stepped-wedge setting using data from two trials. Our goal in this section is to show how RVE changes uncertainty quantification in practice under alternative effect specifications. For each outcome, we fit both an IT model and an ETI model, and we report inference using (i) conventional model-based SEs and (ii) the RVE approach. We present two worked examples to demonstrate both a continuous and a binary outcome. These examples are intended to illustrate the practical implications of the simulation findings rather than to establish substantive conclusions about intervention effectiveness in the two trials. In particular, they show how the choice between model-based and robust variance estimation may alter uncertainty quantification, how ETI specifications typically yield less precise inference than IT specifications, and how inference for the LTE is often substantially more unstable than inference for the TATE. In the illustrative examples, we use the MD-based RVE for the continuous outcome and the MD- and MBN-based RVEs for the binary outcome.

\subsection{Continuous outcome: The Nepal cookstove trial}
The Nepal Cookstove Intervention Trial was a cross-sectional SW-CRT conducted in rural communities in Sarlahi District, Nepal \cite{katz_impact_2020}. The primary objective of the trial was to evaluate whether installation of improved biomass stoves with chimneys, relative to traditional open-fire stoves, improved birth outcomes among pregnant women residing in these communities. Outcomes were identified through ongoing household surveillance. Overall, 51 clusters were randomized to 12 sequences corresponding to the month of intervention delivery, and a total of 
$N=2,379$ live-born infants were enrolled. In our re-analysis, calendar time is modeled categorically by period, rather than as a continuous trend, because the ETI estimand is defined through contrasts across discrete exposure-time periods.

Infant birth weight (in grams) was one of the primary outcomes. Figure~\ref{fig:dataex}(a) reports point estimates and uncertainty summaries for infant birth weight under the IT and ETI specifications in a forest plot, using both conventional model-based SEs and RVE with MD adjustment, combined with an \(I-2\) degrees-of-freedom \(t\)-distribution. Figure~\ref{fig:etivsit} (top panel) illustrates how the treatment effect varies with exposure time.

The IT and ETI models differ in the structural restrictions they impose on the effect trajectory. The IT model assumes a time-invariant treatment effect, so the LTE is identical to the TATE by construction. In contrast, the ETI model allows the effect to evolve with exposure duration and therefore implies a distinct LTE. Under the IT model, the estimated TATE is close to zero and slightly negative ($-2.5$), with SE $45.4$ (MD-adjusted SE $51.3$), corresponding to 95\% CIs of $[-93.8,\,88.7]$ (model-based) and $[-105.7,\,100.7]$ (robust). Under ETI, the TATE is positive ($105.3$) with larger uncertainty (SE $68.5$; MD-adjusted SE $74.5$), yielding wider intervals of $[-32.4,\,242.9]$ (model-based) and $[-44.4,\,254.9]$ (robust).

The practical contribution of RVE is most visible for the long-run estimand under ETI, where inference relies more heavily on information spanning multiple exposure durations and is therefore more sensitive to the working covariance assumptions. The ETI LTE estimate is $125.4$, but it is extremely imprecise under model-based inference (LTE SE $339.4$, 95\% CI $[-556.7,\,807.4]$), because only individuals in the cluster-periods that first transition from control to intervention contribute information to this estimate. Applying RVE changes the uncertainty assessment (LTE robust SE $204.8$, robust 95\% CI $[-286.3,\,537.0]$), though the interval remains wide. Taken together, these results show that (i) the sign of the point estimate is model-dependent (IT vs.\ ETI), and (ii) although RVE may help stabilize the SEs, estimating the LTE remains challenging in practice due to the small number of observations that contribute to its identification. Under both specifications, the resulting intervals remain compatible with no effect in this trial.

This example reflects the main simulation findings for continuous outcomes. Relative to the IT model, the ETI model provides a more flexible treatment-effect specification but yields less precise inference, particularly for the LTE. The comparison between model-based and MD-based robust SEs also shows that uncertainty can change meaningfully once robustness to correlation misspecification is incorporated. In line with the simulation results, the practical value of the robust approach here is not to change the substantive conclusion, but to provide a more reliable uncertainty assessment, especially for ETI-based estimands.

\subsection{Binary outcome: The SMARThealth trial}
The SMARThealth trial was a SW-CRT conducted in rural Andhra Pradesh, in which 18 primary health centre (PHC) were randomized to one of three sequences, dictating the timing of a mobile health intervention. The intervention supported task-sharing between village-based community health workers (ASHAs) and PHC doctors and comprised: (i) household cardiovascular disease (CVD) risk assessment by ASHAs using a tablet-based decision-support tool, (ii) electronic referral and clinical decision support for PHC doctors, and (iii) a tracking system to facilitate follow-up care.\cite{peiris_smarthealth_2019} Clusters crossed over from usual care to the intervention sequentially over three 6-month steps. Independent data collectors screened adults aged $\ge 40$ years to establish a high-CVD-risk cohort. At each 6-month step, an independent random sample (approximately 15\% of the high-risk cohort) was evaluated for outcomes. The primary outcome was the proportion of high-risk individuals meeting a systolic blood pressure (SBP) target of $<140$\,mmHg. Because outcome assessment relied on repeated cross-sectional samples rather than longitudinal follow-up of the same individuals, we treat the resulting data structure as approximately cross-sectional for the purposes of illustrating IT and ETI analyses. Time-varying treatment effects are plausible in SMARThealth, since the intervention operated through a cumulative care pathway rather than a one-time exposure.

\begin{figure}[h!]
    \centering
    \includegraphics[width=\linewidth]{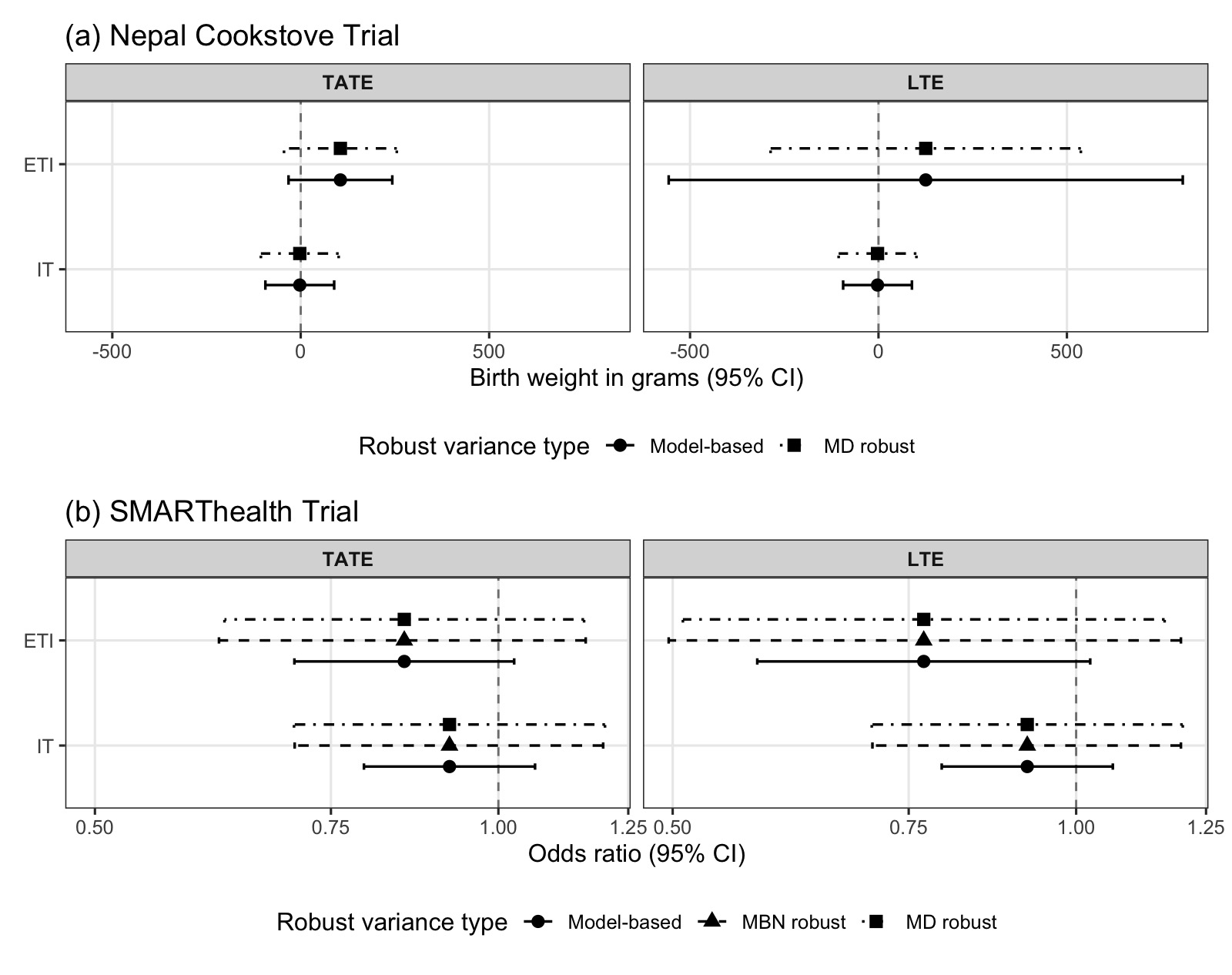}
    \caption{Comparison of IT and ETI estimates under alternative variance estimators. Panel (a) presents TATE and LTE estimates for the Nepal cookstove trial on the continuous outcome scale. Panel (b) presents the corresponding SMARThealth estimates as odds ratios for meeting systolic blood pressure (SBP) targets ($<140$\,mmHg) between two conditions, obtained by exponentiating log-scale coefficients. Points denote point estimates and horizontal bars denote 95\% confidence intervals. Robust intervals are shown using MBN variance estimators.}
    \label{fig:dataex}
\end{figure}

\begin{figure}[h!]
    \centering
    \includegraphics[width=\linewidth]{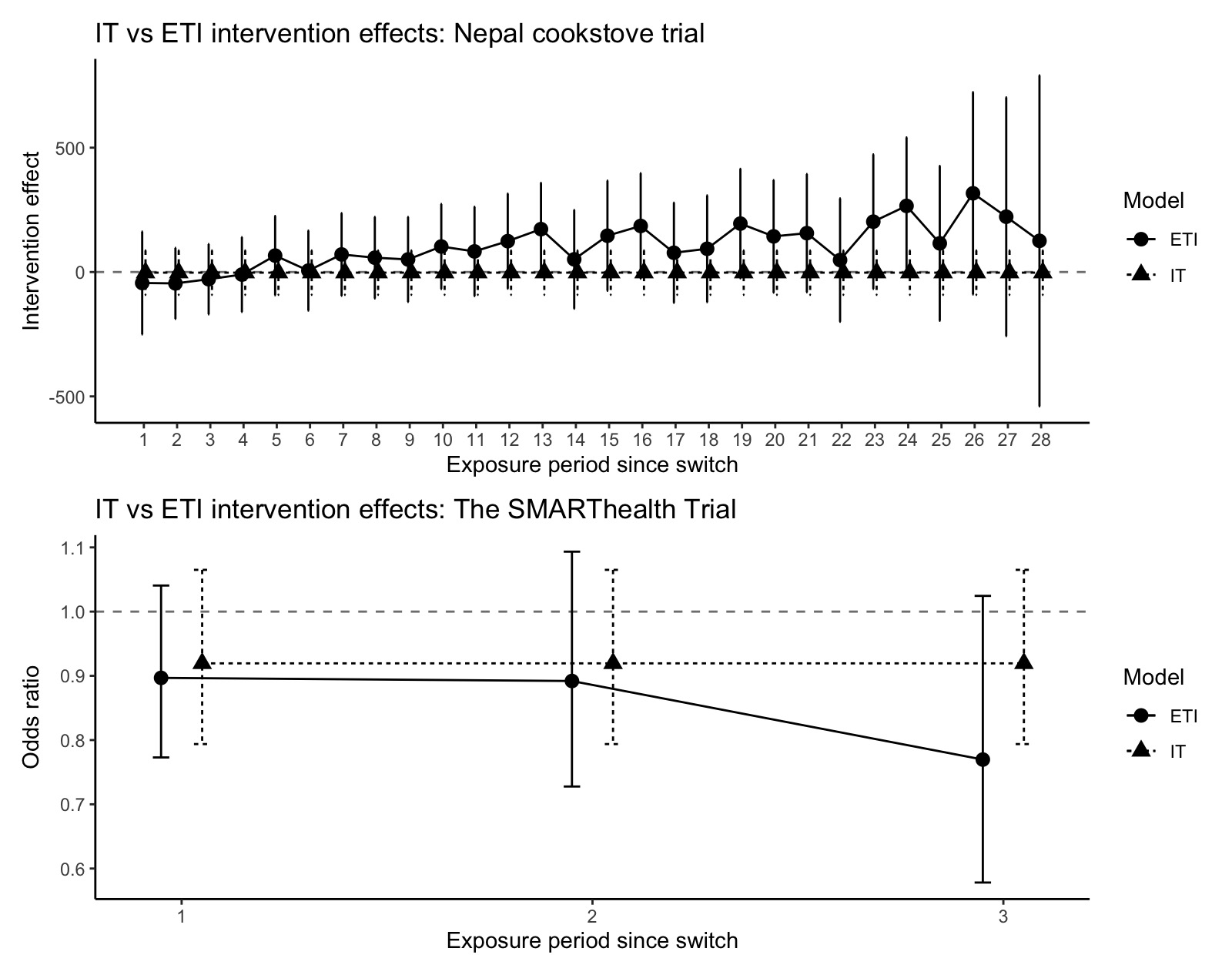}
    \caption{Estimated treatment effects by exposure time under the IT and ETI models for the Nepal cookstove trial and the SMARThealth trial. The IT model assumes a constant treatment effect across exposure time, while the ETI model allows the effect to change with exposure time.}
    \label{fig:etivsit}
\end{figure}

As shown in Figure~\ref{fig:dataex} (b), the IT and ETI models both suggest protective effects for the intervention, but ETI attributes a stronger treatment effect than IT, especially in the long term. Under IT, the estimated TATE is $OR= 0.92$ with 95\% CIs of $[0.79,\,1.07]$ (model-based), $[0.70,\,1.20]$ (MBN) and $[0.70,\,1.20]$ (MD), all of which include the null. Under ETI, the TATE estimate of the treatment effect is even stronger  ($OR=0.85$), and the upper bound slightly exceeds 1, although the 95\% confidence interval is wide:  $[0.70,\,1.03]$ (model-based), $[0.62,\, 1.17]$ (MBN) and $[0.62,\,1.16]$. The contrast is amplified for the long-term effect. Under ETI, the LTE estimate is $OR=0.77$ with even wider 95\% CI: $[0.58, \, 1.02]$ (model-based), $[0.50, 1.19]$ (MBN) and $[0.51, 1.16]$ (MD). As shown in the forest plot, the confidence intervals for LTE tend to be wider than those for TATE, reflecting the greater uncertainty associated with long-term effect estimation. Nevertheless, the point estimates consistently indicate a more protective effect under ETI than under IT, particularly for the LTE. The performance of MD and MBN is similar which is expected given the size of the trial (n=62,254). Figure~\ref{fig:etivsit} (bottom panel) shows that treatment effects can vary across exposure time under both IT and ETI models. Overall, the ETI model is consistently less precise (larger SEs and wider CIs), and in both specifications the (robust) 95\% CIs include the null value, so the differences in magnitudes under IT and ETI do not suggest different conclusions.

\section{Discussion}
This simulation study offers a detailed assessment of RVEs for ETI models in SW-CRTs, extending recent work on model-robust inference for SW-CRTs beyond IT models and continuous outcomes \cite{wang_how_2024,ouyang_maintaining_2024,hughes_robust_2020,thompson_comparison_2021, ford_maintaining_2020}.  Prior studies have shown that, under misspecified random-effects structures, LMMs can yield seriously biased SEs and undercoverage, even when point estimates remain approximately unbiased \cite{wang_how_2024,ouyang_maintaining_2024}.  Our findings reinforce this message in the ETI setting. When the random-effects correlation structure is misspecified, model-based SEs become unreliable and yield confidence intervals with substantial undercoverage, particularly with few clusters. This underscores the need to pair ETI models with appropriately calibrated RVEs rather than relying solely on model-based variances.

Among the RVEs we considered, the bias-corrected MD and MBN estimators consistently outperformed the classic and KC estimators across a wide range of designs, cluster counts, and ICC values. This pattern is consistent with the broader small-sample RVE literature, where higher-order bias corrections and degrees-of-freedom adjustments are essential for reliable coverage \citep{mancl_covariance_2001,wang_covariance_2016,mcneish_modeling_2016, mcneish_effect_2016, morel_small_2003}.  For continuous outcomes, the MD estimator paired with a $t$-distribution provided the most accurate and stable coverage for the TATE across nearly all scenarios. The MBN estimator was also competitive but tended to be mildly conservative when combined with a $t$-distribution. These results mirror the conclusion that approximate jackknife–type RVEs with appropriate degrees-of-freedom corrections often achieve the best balance of bias and coverage for SW-CRTs with small numbers of clusters. \cite{ouyang_maintaining_2024}  

For binary outcomes, the relative performance of RVEs shifted. The MBN estimator under normal distribution offered the most stable coverage across designs, whereas the MD estimator’s coverage deteriorated markedly in sparse-data settings (e.g., 1CPS designs), making MD an unreliable default choice for logistic ETI models. This is consistent with broader experience that MD-type corrections can overinflate variances and produce erratic SEs under extreme sparseness, whereas MBN-type corrections tend to be more stable but conservative. \citep{mancl_covariance_2001,mcneish_effect_2016, morel_small_2003}  In our simulations, the conservatism of the MBN estimator when paired with a $t$-distribution was largely alleviated by using a Normal reference distribution, which brought coverage close to the 95\% nominal level, although coverage remained somewhat conservative when $I \ge 16$ clusters.

Our work also connects directly to the expanding literature on time-varying treatment effects and estimands in SW-CRTs. Several recent contributions have clarified how ETI models target exposure-time–based estimands such as the TATE and LTE, and how these differ from immediate-treatment estimands when treatment effects evolve over exposure or calendar time \cite{maleyeff_assessing_2023,kenny_factors_2025,lee_analysis_2025,hughes_sample_2024}.  Kenny et al. and Hughes et al. have emphasized that power for long-term exposure contrasts is typically much lower than for short-term effects and that design choices (number of clusters, number of periods, and pattern of treatment rollout) strongly influence the information available for TATE- and LTE-type estimands \cite{kenny_factors_2025,hughes_sample_2024}.  Lee et al. further show that misspecifying the treatment-effect time scale (exposure vs. calendar time) can yield severely misleading estimates, including estimators that converge to values with opposite sign from the true time-averaged effect \cite{lee_analysis_2025}. Our study complements these results by demonstrating that, even when the ETI model is correctly specified for the treatment-effect structure, inference for both TATE and especially LTE can fail badly if the variance estimator is not robust and carefully calibrated. 

A recurring and important finding across our simulations was the substantial difficulty of obtaining stable inference for the LTE. Both bias and CI coverage for the LTE were consistently more variable than for the TATE, even under the best-performing RVEs, and poor LTE performance could in turn degrade estimation of the TATE in sparse settings. This aligns with theoretical and empirical work showing that long-term exposure contrasts and late-period point treatment effects are intrinsically low-information and generally require more clusters or longer follow-up to attain reliable precision\cite{maleyeff_assessing_2023,kenny_factors_2025,hughes_sample_2024}. 

Conceptually, the LTE in the ETI models is typically estimated from a relatively small, highly specific subset of the data corresponding to the longest exposure durations. In many SW-CRT designs, especially in 1CPS designs, this subset collapses to a single cluster-period cell. \citep{kenny_factors_2025,watson_grand_2025} Our results highlighted two practical consequences of this design-induced structural sparsity. First, for binary outcomes, estimation of the mean model can become unstable because the LTE is informed by an extremely high-leverage cell. When few or no events are observed in this final exposure cell, the ETI models can produce finite-sample bias, quasi- or complete separation, or highly unstable LTE estimates, with instability propagating to summary estimands such as the TATE. Second, instability can persist even when the fitted model appears well-behaved. In our simulations, the MD estimator's approximate jackknife procedure was highly sensitive to weak support for the longest-exposure indicator, producing inflated SEs and erratic coverage even in settings without obvious zero-cell problems or infinite estimates. By contrast, the MBN estimator remained substantially more stable, suggesting that its bias-correction mechanism is more robust to the structural sparsity induced by the longest-exposure component of the ETI model. When the number of events is small in the longest exposure time period, if scientifically reasonable, one could combine exposure times to estimate the LTE \citep{hughes_sample_2024}

These findings dovetail with recent design work highlighting the vulnerability of 1CPS and other heavily “front-loaded” designs for long-term exposure estimands, especially when outcomes are rare \cite{watson_grand_2025,kenny_factors_2025}.  Even when power calculations suggest adequate precision for a long-term TATE, LTE estimation in ETI models can be practically unstable, with SEs and coverage highly sensitive to the chosen RVE. Investigators whose primary estimand is the LTE should therefore consider designs that reduce sparsity in the longest exposure categories. This may require design choices that would be less important, or even inefficient, if the primary target were the IT effect. For example, one could increase the number of intervention periods to all clusters, extend follow-up with additional late trial periods, or use rollout schemes in which the longest exposure category is represented by multiple cluster-periods rather than a single terminal period. More generally, staircase or incomplete stepped-wedge variants may spread long-exposure information more evenly across the design, rather than leaving LTE precision to depend mainly on variance corrections at the analysis stage. \citep{watson_grand_2025,hemming_analysis_2017}.

\subsection{Limitations}
Although our simulation study was extensive, several limitations bound the scope of our conclusions and suggest directions for future research.

First, we focused on a specific set of random-effects structures and ETI models. Our primary interest was in scenarios where the working correlation structure (e.g., a random intercept or simple cluster-by-period structure) is misspecified, motivated by accumulating evidence that correlation misspecification is both common and consequential in SW-CRTs.  We did not consider more complex structures such as discrete-time decay or random intervention effects, nor did we consider generalized estimating equations with small-sample corrected RVEs in the ETI context. Although existing work suggests that many of our conclusions may extend to these settings, they have not been formally verified for ETI-specific estimands.

Second, our simulations were restricted to cross-sectional SW-CRTs with balanced and complete designs. In practice, investigators frequently encounter incomplete, staircase, or hybrid designs, as well as cohort and open-cohort SW-CRTs in which within-individual correlation introduces additional complexity \cite{grantham_staircase_2024, hughes_current_2015}. Our results, therefore, may not generalize directly to these alternative designs.

Third, we considered only continuous outcomes analyzed via linear mixed models and binary outcomes via logistic mixed models. For binary outcomes, although data were generated under the ETI framework, we considered only a flat exposure-time effect profile with $\delta_1=\cdots=\delta_{J-1}$. We did not evaluate settings in which the true binary treatment effect changed across exposure periods. We did not investigate count outcomes (e.g., Poisson or negative binomial models) or alternative link functions (e.g., log or complementary log–log). For these outcome types, estimation difficulties (non-convergence, quasi-separation) are often more severe with few clusters, and both model-based and robust variance estimators can behave poorly. Consequently, the robustness of MD and MBN estimators for ETI models with non-Gaussian, non-binary outcomes cannot be assumed and warrants a dedicated investigation. We also emphasize that in settings with binary outcomes, differences across GLMM-based estimands can arise not only from model misspecification but also from the intrinsic non-collapsibility of the odds ratio. As a result, even when the fixed-effects component is correctly specified, conditional effects from mixed models may not coincide with marginal effects, and such discrepancies may become more pronounced as the magnitude of between-cluster heterogeneity increases. Our findings, therefore, should be interpreted in light of this property. Part of the observed divergence between IT- and ETI-based estimands (and their corresponding LTE summaries) may reflect non-collapsibility rather than true differences in exposure-time dynamics. Importantly, this feature does not undermine the relevance of our evaluation. Instead, it highlights that RVE-based inference is being assessed under realistic conditions faced by analysts of SW-CRTs with binary outcomes, where non-collapsibility is unavoidable and may meaningfully affect finite-sample behavior.

Finally, we restricted attention to ETI models that treat exposure-time effects as categorical, which is the most flexible but often the least efficient specification. Alternatively, more parsimonious specifications, for example, delayed-constant effects, piecewise-linear trends, or spline-based exposure-time trends, have been proposed to improve efficiency and power. Our results do not directly address how MD or MBN estimators perform under such structured ETI models, nor how best to combine flexible ETI models with regularization or shrinkage to stabilize LTE estimation. This remains an important area for future research.

\subsection{Conclusion}
In conclusion, our study provides outcome-specific, practically oriented guidance for analyzing SW-CRTs with ETI models under correlation misspecification and small-sample conditions. For the TATE with continuous outcomes, we recommend the MD estimator paired with a $t$-distribution with $I-2$ degrees of freedom as a default choice, given its favorable bias–variance trade-off and stable coverage across a wide range of designs. For the TATE with binary outcomes, we recommend the MBN estimator as the most reliable option, preferably paired with a Normal reference distribution to mitigate small-sample conservatism. These recommendations are broadly consistent with earlier work suggesting that approximate jackknife RVEs with small-sample corrections (MD-type) and bias-inflated sandwich RVEs (MBN-type) provide the most reliable coverage among competing RVEs in clustered data with few clusters. We strongly discourage use of the MD estimator for binary outcomes in settings where data sparsity is likely, particularly in 1CPS or other designs in which the longest exposure period is supported by only a single or very small number of informative cluster-period cells.

More broadly, our results reinforce that in SW-CRTs with time-varying treatment effects, valid inference depends jointly on (i) careful definition of the estimand (e.g., immediate treatment effect, TATE, LTE), (ii) appropriate specification of the time scale and treatment-effect structure (exposure vs. calendar time, or both), and (iii) a variance estimator that remains robust under realistic small-sample regimes.  Importantly, we do not view time-varying treatment effects, or ETI models more generally, as universally applicable or as a default analysis that should always be fitted first and then simplified if no pattern is detected. Rather, the choice between an IT model and a more complex ETI model should be made a priori, based on substantive knowledge about how the intervention is expected to exert its effect, much as investigators prespecify whether an interrupted time series analysis is expected to show a step change, a slope change, or both. In practical terms, ETI models are most appropriate when the intervention effect is plausibly expected to evolve with time since implementation. For example, because uptake is gradual, implementation fidelity improves over time, effects accumulate biologically or behaviorally, or benefits emerge only after a lag. By contrast, an IT model may be sufficient when the intervention is expected to produce a prompt and approximately stable effect once introduced. Investigators should therefore prespecify not only the primary estimand but also the primary treatment-effect structure, drawing on prior evidence, mechanistic understanding, and the expected timing of outcome response. When uncertainty remains, alternative ETI specifications may be considered as secondary or sensitivity analyses rather than selected post hoc from the observed data. This is especially important in small-sample SW-CRTs, where the available information may not well support more complex exposure-time models. Given the persistent instability we observed for LTE estimation, even under the best-performing RVEs and especially for binary outcomes, we recommend that researchers prioritize the TATE as their primary estimand wherever possible, treating LTE-type summaries as secondary or exploratory. This recommendation aligns with recent design and power work, highlighting the high information requirements for reliable long-term exposure contrasts. Future research should focus on developing novel estimators and design strategies that explicitly target stable long-term effects in small-sample SW-CRTs, building on recent advances in ETI-based estimands, model-robust inference, and staggered-implementation trial design.

\section*{Acknowledgments}
Research in this article was partially supported by a Patient-Centered Outcomes Research Institute Award\textsuperscript{\textregistered} (PCORI\textsuperscript{\textregistered} Award ME-2022C2-27676) and NIH grant AI029168. The statements presented are solely the responsibility of the authors and do not necessarily represent the official views of PCORI\textsuperscript{\textregistered}, its Board of Governors, or the Methodology Committee. We appreciate Dr. Joanne Katz for sharing the Nepal cookstove trial data and reviewing the illustration example section

\section*{Conflict of interest}
The author declares no potential conflict of interest.

\section*{Supporting information}
Additional supporting information, including Web Appendices A, Web Tables, may be found online in the supporting information tab for this article.
The code for calculating robust variance estimation is available at: https://github.com/jphughes9/vcovCRglmer

\bibliography{SWDdecay}

\end{document}